\documentclass[superscriptaddress,twocolumn]{revtex4}
\usepackage{amsmath,lscape,epsfig}

\def\beq{\begin{equation}}
\def\eeq{\end{equation}}
\def\beqa{\begin{eqnarray}}
\def\eeqa{\end{eqnarray}}
\def\ban{\begin{eqnarray*}}
\def\ean{\end{eqnarray*}}
\def\bi{\begin{itemize}}
\def\ei{\end{itemize}}

\newcommand{\kv}{\mathbf{k}}
\newcommand{\qv}{\mathbf{q}}
\newcommand{\rv}{\mathbf{r}}

\newcommand{\s}{\rm s}

\newcommand{\mF}{\mathcal{F}}
\newcommand{\mH}{\mathcal{H}}
\newcommand{\mU}{\mathcal{U}}

\newcommand{\pprime}{\prime\prime}

\newcommand{\be}{\begin{eqnarray}}
\newcommand{\ee}{\end{eqnarray}}

\begin{document}

\title{
Cluster formation in compact stars:
relativistic versus Skyrme nuclear models}

\author{C. Ducoin}
\affiliation{Istituto Nazionale di Fisica Nucleare, Sezione di Catania, Via Santa Sofia 64, I-95123 Catania, Italy}
\affiliation{LPC (IN2P3-CNRS/Ensicaen et Universit\'{e}), F-14050 Caen C\'edex, France}
\author{C. Provid\^encia}
\affiliation{Centro de F\'{\i}sica Computacional, Department of  Physics,
University of Coimbra,  P-3004 - 516 Coimbra, Portugal}
\author{A. M. Santos}
\affiliation{Centro de F\'{\i}sica Computacional, Department of  Physics,
University of Coimbra,  P-3004 - 516 Coimbra, Portugal}
\author{L. Brito}
\affiliation{Centro de F\'{\i}sica Computacional, Department of  Physics,
University of Coimbra,  P-3004 - 516 Coimbra, Portugal}
\author{Ph. Chomaz}
\affiliation{GANIL (DSM-CEA/IN2P3-CNRS), B.P. 5027, F-14076 Caen
C\'edex 5, France}

\begin{abstract}
We present various properties of nuclear and compact-star matter,
comparing the predictions from two kinds of phenomenological approaches:
relativistic models (both with constant and density-dependent couplings)
and non-relativistic Skyrme-type interactions.
We mainly focus on the liquid-gas instabilities that occur at sub-saturation densities,
leading to the decomposition of the homogeneous matter into a clusterized phase.
Such study is related to the description of neutron-star crust (at zero temperature)
and of supernova dynamics (at finite temperature).
\end{abstract}

\maketitle

\vspace{0.0cm} PACS number(s):
{21.65.+f,24.10.Jv,21.30.-x,21.60.-n}
\vspace{0.50cm}

\section{Introduction}


The knowledge of the equation of state (EOS) of nuclear matter
under exotic conditions is essential for our understanding of the nuclear force 
and for astrophysical applications.
This implies high isospin asymmetries, finite temperatures, and a wide density range
(both for subsaturation and suprasaturation densities).
The next generation of observational and experimental data 
is expected to bring new constraints in order to refine the theoretical models:
for instance, the forthcoming radioactive-ion-beam facilities (such as FAIR@GSI and SPIRAL2@GANIL)
will allow the investigation of the isospin degree of freedom in nuclear structure and dynamics.

The present work is dedicated to the predictions of different effective nuclear models.
It is mainly focused on the liquid-gas instabilities
present in nuclear and stellar matter at sub-saturation density.
These instabilities are directly related to the bulk EOS.
They are used to explain the multifragmentation phenomenon 
occurring in collisions around the Fermi energy~\cite{multi}:
in the spinodal decomposition scenario, 
fragment formation is induced by the fast development of spinodal instabilities
in the low-density expanding matter formed just after the collision~\cite{chomaz-PhysRep}.
Finite-size liquid-gas instabilities are also important for compact-star physics:
matter non-homogeneities in the (hot) core of type II supernovae 
is expected to affect the dynamics of the explosion,
and the crust of (cold) neutron stars contains a 
non-homogeneous phase commonly named {\em pasta phase}~\cite{pethick,horo,maruyama05}.
It should be noticed that the study of liquid-gas instabilities
is complementary to the equilibrium approaches which are also used 
to describe the clusterized stellar matter:
namely (at very low density) the virial equation of state ~\cite{hor06},
and (at higher density) the calculation of the pasta phases as the ground state shaped by the competition between Coulomb repulsion and surface tension.
Although nuclear equilibrium is expected to be reached in most stellar conditions,
the spinodal-instability properties should help
to understand the physics of  compact stars in the following ways:
(i) giving an estimation of clusterized-matter properties,
such as cluster size and composition;
(ii) showing the minimal region where the equilibrated matter must be formed of clusters; 
(iii) possibly, playing a direct role in cluster formation for specific situations involving very short time scales (as may happen during a supernova explosion).


In this paper, we compare predictions from two kinds of models 
based on phenomenological density functionals:
relativistic and non-relativistic.
Both are commonly used to describe asymmetric nuclear matter,
in the framework of exotic nuclei as well as compact stars.
However, it is well-known that, although all give a quite good description of stable nuclei
(consistently with the constraints included in the fitting procedures),
they present different behaviours as soon as exotic conditions are reached,
especially in the isovector channel.
Our scope is to explore the impact of these different behaviors
on quantities of interest for compact-star physics, 
such as the clusterization properties and the matter composition at $\beta$-equilibrium.

As a non-relativistic approach, 
we use the effective density dependent Skyrme-type interaction~\cite{original,brink,jirina07}.
The simple form of the Skyrme functional makes it an attractive model 
for the description of both nuclei and compact-star matter.
It was originally intended to describe nuclear properties through the mass table,
and the older parametrizations 
only include in their fits constraints from magic-nucleus properties along the stability line.
Trying to give a reliable description of exotic nuclei and stellar matter,
the modern Skyrme parametrizations 
also include in their fitting procedures results from microscopic calculations of neutron-rich matter.
The Skyrme-Lyon (SLy) forces for instance have been used in studies of neutron-star crust~\cite{Douchin-PLB485, Douchin-NPA665}. 
Such parametrizations are among the 27 forces which were not ruled out 
for unfit neutron-star properties in the extensive study by J.R.~Stone et al.~\cite{stone}, 
where 87 Skyrme parametrizations were checked.

In contrast with the Skyrme approach, 
the relativistic nuclear models are, by construction, causal
and can thus be applied to a wider region of the compact stars
(as long as matter is supposed to be in an hadronic phase).
Relativistic Mean-Field models (RMF) have been used to describe the EOS of compact
stars~\cite{prak97,glen00,compact1}, both cold and warm. 
The Density Dependent Hadronic models (DDH)
are an alternative approach for the description of nuclear matter and finite nuclei~\cite{fuchs}.
In DDH, the non-linear self-interactions of the mesons occurring in constant coupling models
are substituted by density-dependent meson-nucleon coupling parameters,
motivated by Dirac-Brueckner calculations of nuclear matter.
Such models are found to behave more closely to the non-relativistic ones.


Relativistic and Skyrme approaches have been compared
from the formal point of view in a recent work~\cite{bcmp07},
where a low-density expansion of the RMF and DDH models have been used
in order to directly compare the different density functionals.
oth kinds of models have been used separately in several previous works
for the study of spinodal instabilities in nuclear and compact-star matter,
at zero and finite temperature:
see for instance Refs.~\cite{inst04, inst06, umodes06,umodes06a,umodes08} for relativistic models,
and Refs.~\cite{JMPC-PRC67, CD-A2, CD-A3} for Skyrme models.
The same qualitative features are reproduced 
(general shape of the instability regions, isospin-distillation property of the phase separation).
The scope of the present paper is then to have a direct look at the quantitative differences
between relativistic and Skyrme-model predictions.
We wish to investigate the extent to which the 
different temperature and  isospin dependences of the nuclear 
EOS can affect the neutron-star properties
and determine the sensitive features that have to be constrained.



In section II we briefly review the relativistic and Skyrme models used in the present work. 
In section III, we present the Vlasov formalism 
that we use to address the dynamic instabilities, in both frameworks.
Nuclear-matter properties are discussed for the different models in section IV, 
where we present the nuclear EOS (in isoscalar and isovector channel) 
as well as the spinodal instabilities (in both thermodynamical and dynamical frameworks).
Properties of stellar matter, 
including homogeneous $\beta$-equilibrium matter 
and instabilities against clusterization, are discussed in section V. 
In the last section we draw the main conclusions from our work.


\section{\protect\smallskip  Effective Nuclear Models\label{SEC:models}}

In the sequel we will give a short presentation of the models discussed in the present paper. 
All expressions are given in units $\hbar=c=1$.
We will consider first density functionals based on  Skyrme forces, then the RMF and DDH models.
The nuclear matter saturation properties obtained with all models used in the present work
are reported in Table \ref{tab:properties}.

\begin{table*}[hbt]
  \centering
\caption{Nuclear matter properties of the Skyrme and relativistic models used in the present work.}
  \begin{tabular}{lclcccrr}
\hline
    Model & $B/A$ & $\rho_0$ & $K$ &  $m^*/m$&  $a_{\s}(\rho_0)$ &$L(\rho_0)$&
$K_{\rm{sym}}(\rho_0)$    \\
          & (MeV)&(fm$^{-3}$)& (MeV)& &(MeV)& (MeV) & (MeV)\\
\hline
SIII~\cite{SIII}      &15.9 &0.145 &356&0.76&    28.2 &9.9 & -394 \\
SGII~\cite{SGII}      &15.6 &0.159 &215&0.79&    26.9 & 37.6 & -146\\
SLy230a~\cite{Slya}   &16.0 &0.16  &230&0.70&    32.0 & 44.3 & -98\\
NRAPR~\cite{NRAPR}    &15.9 &0.16  &226&0.70&    32.8 & 59.6 & -123\\
LNS~\cite{LNS}        &15.3 &0.175 &211&0.83&    33.4 & 61.5 & -127\\
\hline
NL3~\cite{nl3}        &16.3 &0.148 &272 &0.60&    37.4 & 118.3& 101\\
NL$\delta$~\cite{liu} &16.0 &0.160 &240 &0.75&    30.5 & 102.7& 127\\
TW~\cite{tw}          &16.3 &0.153 &240 &0.56&    32.0 & 55.3& -125\\
DD-ME2~\cite{ring02}  &16.1 &0.152 &251&0.57&    32.3 & 51.7 & -88\\
 DDH$\delta$~\cite{gaitanos}&16.3&0.153& 240&0.56&25.1 & 48.6& 81\\
\hline
  \end{tabular}
  \label{tab:properties}
\end{table*}


\subsection{\protect\smallskip Skyrme functional}

The local Skyrme interaction~\cite{brink}
allows to introduce an energy density $\mathcal{H}{(\mathbf{r})}$ so that
the total energy for a system of nucleons in a Slater determinant 
$\mid \psi>$ reads :
\begin{equation}
\langle \psi | \hat{H}| \psi \rangle=\int {\mH(\rv)d^3r}
\label{EQ:meanH}
\;,
\end{equation}
where $\mH(\rv)$ is the Skyrme energy-density functional.

In the case of homogeneous, spin-saturated matter with no
Coulomb interaction, the Skyrme energy-density functional~\cite{Slya} reduces to four terms:
\be
\mathcal{H}^b=\mathcal{K}+\mathcal{H}_{0}+\mathcal{H}_{3}+\mathcal{H}_{\rm{eff}} 
\label{EQ:mathH}
\ee
where the label $b$ (bulk) is used to mark the thermodynamic framework.
In this expression, $\mathcal{K}$ is the kinetic-energy term, 
$\mathcal{H}_{0}$ a density-independent two-body term, $\mathcal{H}_{3}$ 
a density-dependent term, and $\mathcal{H}_{\rm{eff}}$ a momentum-dependent term:
\be
\mathcal{K}&=&\frac{\tau}{2m}\\ 
\mathcal{H}_{0} &=&C_{0}\rho ^{2}+D_{0}\rho _{3}^{2}\\
\mathcal{H}_{3} &=&C_{3}\rho ^{\sigma +2}+D_{3}\rho ^{\sigma }\rho _{3}^{2}\\
\mathcal{H}_{\rm{eff}} &=&C_{\rm{eff}}\rho \tau +D_{\rm{eff}}\rho _{3}\tau _{3}
\;.
\ee
We have introduced the isoscalar and isovector particle densities, $\rho$ and $\rho_3$, 
as well as kinetic densities, $\tau$ and $\tau_3$: 
\begin{equation}
\begin{array}{lll}
\rho =\rho _{n}+\rho _{p} & ; & \tau =\tau _{n}+\tau _{p} \\ 
\rho _{3}=\rho _{n}-\rho _{p} & ; & \tau _{3}=\tau_{n}-\tau _{p}
\end{array}
\end{equation}
where, denoting $i$ the third component of the isospin
($n$ for neutrons and $p$ for protons),
the kinetic densities are defined by $\tau_i=\langle \hat{k}^2\rangle_i$.
The coefficients $C$ and $D$, associated respectively 
with the symmetry and asymmetry contributions, are linear combinations of the traditional Skyrme parameters:
\be
\begin{array}{ll}
C_{0}&= \ \ 3t_{0}/8 \\ 
D_{0}&=- t_{0}(2x_{0}+1)/8 \\ 
C_{3}&= \ \ t_{3}/16 \\ 
D_{3}&=-t_{3}(2x_{3}+1)/48 \\ 
C_{\rm{eff}}&= \ \ [3t_{1}+t_{2}(4x_{2}+5)]/16 \\ 
D_{\rm{eff}}&= \ \ [t_{2}(2x_{2}+1)-t_{1}(2x_{1}+1)]/16
\label{EQ:SkyrmeCoef}
\end{array}
\;.
\ee
In the mean-field approach, the individual particle level is derived from this functional.
For each particle species, it is given by:
\be
\label{EQ:ind-levels}
\hat h_i^b
&=&m_i+\frac{\partial\mH^b}{\partial\rho_i} +\frac{\partial\mH^b}{\partial\tau_i} \hat k_i^2 \\
&=&m_i+U_i+\frac{1}{2m^*_i}\hat k_i^2 \;.
\ee
where we have included the nuclear mass energy $m$.
The kinetic energy is expressed in the non-relativistic limit,
in terms of an effective mass $m^*_i$ defined by:
\be
\frac{1}{2m^*_i}=\frac{1}{2m_i}+\frac{\partial \mH_{\rm{eff}}}{\partial\tau_i}
\;.
\ee

The chemical potentials $\mu_i$ are such that the Fermi-Dirac occupation number is:
\be
n_i(k)=\left[1+e^{\beta(\frac{k^2}{2m^*_i}+U_i-\mu_i)}\right]^{-1}
\;;
\ee
we can also define a chemical potential $\mu^t_i=\mu_i+m_i$ including the mass energy, such that:
\be
n_i(k)=\left[1+e^{\beta(m_i+\frac{k^2}{2m^*_i}+U_i-\mu^t_i)}\right]^{-1}
\;.
\ee

In this work, we will use conventional and modern Skyrme interactions. 
The earlier parametrizations, such as SIII~\cite{SIII} and SGII~\cite{SGII},
have been established by fitting the properties of stable nuclei 
(such as radii and ground-state energy).
They are thus in principle poorly adapted to a description of neutron-rich matter.
It is indeed found that SIII presents an irrealistic behavior in the isovector channel;
furthermore it has a too high incompressibility at saturation. 
As a result, this interaction will present an atypical behavior all through the following study.
SGII, however, for which spin properties have also been used as constraints, 
presents a more reasonable evolution in the isovector channel.
In particular, it has been shown to reproduce isospin effects in giant dipole resonances~\cite{SGII}.
Among the modern Skyrme-type parametrizations, 
we have chosen to use one of the Skyrme-Lyon forces (SLy230a~\cite{Slya}),
as well as the NRAPR~\cite{NRAPR} and LNS~\cite{LNS} parametrizations.
All these recent forces include in their fitting procedure results from microscopic calculations.
SLy230a uses the pure-neutron matter equation of state UV14+UVII by R.B. Wiringa et al~\cite{Wiringa-PRC38}.
NRAPR (Non-Relativistic APR) stands for the Skyrme interaction parameters obtained 
from a fitting to the APR equation of state (Akmal-Pandharipande-Ravenhall, Ref.~\cite{akmal}).
LNS is based on Brueckner-Hartree-Fock calculations of infinite nuclear matter at different values of isospin asymmetry. 
Such constrains from microscopic calculations are intended 
to control the behavior of the resulting effective force 
far from saturation and up to high isospin asymmetry.


\subsection{\protect\smallskip Relativistic approaches}

In the present paper, we will consider two kinds of relativistic effective approaches:
RMF models, which have constant coupling parameters described 
by the Lagrangian density of non-linear Walecka models (NLWM),
and DDH models with density-dependent coupling parameters.
In each case, we consider models including or not the $\delta$-meson, 
which have been introduced to include in the isovector channel the same
symmetry existing already in the isoscalar channel with the meson pair $(\sigma,\omega)$
responsible for saturation in RMF models~\cite{liu}.
The presence of the $\delta$-meson softens the symmetry energy at subsaturation densities 
and hardens it above saturation density.
The RMF parametrizations we use are NL3~\cite{nl3} and NL$\delta$~\cite{liu};
the DDH ones are TW~\cite{tw}, DD-ME2~\cite{ring02} and DDH$\delta$~\cite{gaitanos}.
Only NL$\delta$ and DDH$\delta$ include the $\delta$-meson.

The relativistic approach is based on a lagrangian density given by:
\begin{equation}
\mathcal{L}=\sum_{i=p,n}\mathcal{L}_{i}\mathcal{\,+L}_{{\sigma }}%
\mathcal{+L}_{{\omega }}\mathcal{+L}_{{\rho}}+ {\cal{L}}_\delta\;.
\label{lag}
\end{equation}
The nucleon Lagrangians read:
\begin{equation}
\mathcal{L}_{i}=\bar{\psi}_{i}\left[ \gamma _{\mu }iD^{\mu
}-\mathcal{M}^{*}\right] \psi _{i}\;,  \label{lagnucl}
\end{equation}
with 
\begin{eqnarray}
iD^{\mu } &=&i\partial ^{\mu }-\Gamma_{v}V^{\mu }-\frac{\Gamma_{\rho }}{2}{\vec{\tau}}%
\cdot \vec{b}^{\mu }  \label{Dmu} \\
\mathcal{M}^{*} &=&m-\Gamma_{s}\phi -\Gamma_{\delta }{\vec{\tau}}\cdot
\vec{\delta}\;, \label{Mstar}
\end{eqnarray}
where $\vec{\tau}$ is the isospin operator. We use the vector symbol to
designate a vector in isospin space. 

The isoscalar part is associated with the scalar sigma ($\sigma$) field $\phi$ 
and the vector omega ($\omega $) field $V_{\mu }$, 
while the isospin dependence comes from the isovector-scalar delta ($\delta $) field $\delta ^{i}$ 
and the isovector-vector rho ($\rho $) field $b_{\mu }^{i}$
(where $\mu$ is a space-time index and $i$ an isospin-direction index). 
The associated Lagrangians are:
\begin{eqnarray*}
\mathcal{L}_{{\sigma }} &=&+\frac{1}{2}\left( \partial _{\mu }\phi \partial %
^{\mu }\phi -m_{s}^{2}\phi ^{2}\right)-\frac{1}{3!}\kappa \phi ^{3}-\frac{1}{4!}%
\lambda \phi ^{4}  \\
\mathcal{L}_{{\omega }} &=&-\frac{1}{4}\Omega _{\mu \nu }\Omega ^{\mu \nu }+%
\frac{1}{2}m_{v}^{2}V_{\mu }V^{\mu }\\
\mathcal{L}_{{\delta }} &=&+\frac{1}{2}(\partial _{\mu }\vec{\delta}\partial %
^{\mu }\vec{\delta}-m_{\delta }^{2}{\vec{\delta}}^{2}\,)\\
\mathcal{L}_{{\rho }} &=&-\frac{1}{4}\vec{B}_{\mu \nu }\cdot
\vec{B}^{\mu \nu }+\frac{1}{2}m_{\rho }^{2}\vec{b}_{\mu }\cdot
\vec{b}^{\mu }\;,
\end{eqnarray*}
where $\Omega _{\mu \nu }=\partial _{\mu }V_{\nu }-\partial _{\nu}V_{\mu }$, 
$\vec{B}_{\mu \nu }=\partial _{\mu }\vec{b}_{\nu }-\partial _{\nu }\vec{b}%
_{\mu }-\Gamma_{\rho }(\vec{b}_{\mu }\times \vec{b}_{\nu })$, and $\Gamma_{j}$ and $%
m_{j}$ are respectively the coupling parameters of the mesons
$j=s,v,\delta,\rho $ with the nucleons and their masses.
The self-interacting terms for the $\sigma$-meson are included
only for  the NL3 and NL$\delta$ parametrizations, 
$\kappa$ and $\lambda$ denoting the corresponding coupling constants.

The density-dependent coupling parameters
$\Gamma_{s}$, $\Gamma_v$ and $\Gamma_{\rho}$,  are adjusted
in order to reproduce some of the nuclear matter bulk properties,
using the following parametrization:
\begin{equation}
\Gamma_i(\rho)=\Gamma_i(\rho_{sat})f_i(x)\;, \quad i=s,v
\label{paratw1}
\end{equation}
with
\begin{equation}
f_i(x)=a_i \frac{1+b_i(x+d_i)^2}{1+c_i(x+d_i)^2}\;,
\end{equation}
where $x=\rho/\rho_{sat}$ and
\begin{equation}
\Gamma_{\rho}(\rho)=\Gamma_{\rho}(\rho_{sat})
\exp[-a_{\rho}(x-1)]\;. \label{paratw2}
\end{equation}
The values of the parameters $m_i$, $\Gamma_i$, $a_i$, $b_i$,
$c_i$ and $d_i$, $i=s,v,\rho$ for TW and DD-ME2
are respectively given in~\cite{tw} and~\cite{ring02} and for DDH$\delta$ 
in~\cite{gaitanos,inst04}. In this last case the parametrization for the $\delta$
and $\rho$ coupling parameters is also given by (\ref{paratw1}) with
$$f_i(x)=a_i \exp[-b_i(x-1)]-c_i(x-d_i)\;, \quad i=\rho,\,\delta.$$
The $\Gamma_i$ coupling parameters 
are replaced by the  $g_i$ coupling constants in the NL3 and NL$\delta$ models.

\section{\protect\smallskip The Vlasov Formalism\label{SEC:vlasov}}

In this paper, we study the dynamic spinodal instabilities
as unstable density fluctuation modes obtained in the Vlasov framework.
The present section gives a short review of the Vlasov formalism 
already introduced in Refs.~\cite{npp91,stable-modes05,umodes06}
and give the resulting expressions for Skyrme and relativistic models.
For simplicity, we consider here nuclear matter where the proton electric charge is
neutralized in average by a uniform background. 
We thus neglect the electron degree of freedom existing in star matter, 
which is model-independent and has only a  perturbative effect.


\subsection{\protect\smallskip Brief review}
To describe the time evolution of the nuclear system, 
we introduce the one-body phase-space distribution function in isospin space: 
$f(\rv,\kv,t)=\mbox{diag}\left(f_p,f_n\right)$,
and the corresponding one-body Hamiltonian
$h=\mbox{diag}\left(h_{p},h_{n}\right)$.
The time evolution of the distribution function is described by the Vlasov equation:
\begin{equation}
\frac{\partial f_i}{\partial t} +\{f_i,h_i\}=0, \qquad \;
i=p,\,n, \label{vlasov1}
\end{equation}
where $\{,\}$ denotes the Poisson brackets.
At zero temperature, the state which minimises the energy of asymmetric nuclear matter 
is characterised by the Fermi momenta
$k_{Fi}$, $i=p,n$, and is described by the
distribution function 
$f_0(\kv) =
\mbox{diag}\left[\Theta(k_{Fp}^2-k^2),\Theta(k_{Fn}^2-k^2)\right]$.
In order to describe small oscillations around the equilibrium state,
we take for the distribution functions $f_i\,=\, f_{0i} + \delta f_i\;$ 
and introduce a generating function~\cite{npp91} 
$S(\rv,\kv,t)=\mbox{diag}\left( S_p, \, S_n\right)$  
defined in isospin space such that $\delta f_i \,=\, \{S_i,f_{0i}\}$.
In terms of the generating function, the linearised Vlasov equations
for $\delta f_i$ are equivalent to the following time evolution equations:
\begin{equation}
  \label{eq:deltafe}
 \frac{\partial S_i}{\partial t} + \{S_i,h_{0i}\} =(\delta h_i)_F,
\end{equation}
where $(\delta h_i)_F$ is the mean-field variation at Fermi level,
which depends on the considered nuclear model.

We will consider the longitudinal fluctuations such that:
\beq
\label{ansatz}
\left(\begin{array}{lllll}S_i;& \delta \rho_i;& \delta h_i \end{array} \right)
=
\left(\begin{array}{lllll}S_{\omega,i}(x);& \delta \rho_{\omega,i} ;& \delta h_{\omega,i} \end{array}\right)
e^{i(\qv\cdot\rv-\omega t)}
\eeq
where $x=\cos(\kv,\qv)$.
The longitudinal normal modes are obtained substituting the ansatz~(\ref{ansatz}) 
in the linearised equations of motion. 
The dispersion relation takes the form:
\begin{equation}
\left(\begin{array}{cc}
1+F^{pp}L_p & F^{pn}\, L_p\\
F^{np} \, L_n &1+F^{nn}\, L_n\\
\end{array}\right)
\left(\begin{array}{c}
A_{\omega p}\\
A_{\omega n}\\
\end{array}\right)
=0,
\label{eq:rd}
\end{equation}
where $L_i=L(s_i)=2-s_i \ln \left[{(s_i+1)}/{(s_i-1)}\right]$ is the Lindhard function,
with $s_i=\omega/(qv_{Fi})$ in terms of the Fermi velocity $v_{Fi}=\partial\epsilon_{Fi}/\partial k_{Fi}$.

The amplitudes $A_{\omega i}=\int_{-1}^1 x\, S_{\omega i}(x)\, dx$ are
 related to the transition densities by
$$\delta\rho_i=\frac{\omega\, N_{0i}}{2\, s_i}A_{\omega i},$$
where $N_{0i}$ is the density of states at the Fermi surface.
With all models, the coefficients $F^{ij}$ appearing in Eq.~(\ref{eq:rd}) 
can be expressed in terms of two quantities related to the nuclear residual interaction,
$\mU_{ij}^{(1)}$ and $\mU_{ij}^{(2)}$ defined by:
\be 
\label{EQ:dhi-drhoj}
(\delta h_i)_F
&=&\sum_j {\left[\mU_{ij}^{(1)}+x\,\mU_{ij}^{(2)}\right] \delta\rho_j}\;.
\ee
In this expression, we separate the $x$-dependent contribution of the residual interaction 
($x\,\mU_{ij}^{(2)}$) from the x-independent one ($\mU_{ij}^{(1)}$).
On the other hand, Eq.~(\ref{eq:deltafe}) provides the following relation:
\be
\label{EQ:dhi-croch}
(\delta h_i)_F &=& -i\omega S_i(x)  \left[ 1- x/s_i \right]\,.
\ee
From Eqs.~(\ref{EQ:dhi-drhoj}) and~(\ref{EQ:dhi-croch}), we obtain the set of Vlasov equations:
\be
\label{eq:vlasov-drho}
\frac{2}{N_{0i}}\delta\rho_i+ L_i\sum_j {\left[\mU_{ij}^{(1)}+s_i\mU_{ij}^{(2)}\right] \delta\rho_j}=0\,,
\ee
which is equivalent to Eq.~(\ref{eq:rd}) with the following identification:
\be
F^{ij}&=&\frac{N_{0j}}{2}\frac{s_i}{s_j}\left(\mU_{ij}^{(1)}+s_i\,\mU_{ij}^{(2)}\right)\;.
\ee
The model-dependence of the Vlasov equations 
is then contained in the coefficents $\mU_{ij}^{(1,2)}$ that we have introduced:
next, more details will be given for both Skyrme and relativistic models.

Let us note that the instabilities of the system are determined from the imaginary frequencies
which satisfy the dispersion relation~\cite{stable-modes05}.
The finite-size instability region is the envelope of all the dynamical spinodals 
corresponding to different values of the transfered momentum $q$.
For simplicity, through out this work we will identify 
this domain with the dynamical spinodal for $q=80$~MeV,
which is a very good approximation for all the models under study.


\subsection{\protect\smallskip Vlasov approach with Skyrme models}

We identify three contributions to the mean-field variation $\delta h_i$:
the bulk ($b$), surface ($\nabla$) and Coulomb ($c$) terms, such that
\be
\delta h_i&=&\delta h_i^b+\delta h_i^{\nabla}+\delta h_i^c\;.
\ee

The bulk term is:
\be
\delta h_i^b
&=&\delta\left[ U_i+\frac{k^2}{2m^*_i}  \right]\\
&=&\sum_j \left[\frac{\partial^2\mH}{\partial\rho_j\partial\rho_i}
+\frac{\partial^2\mH}{\partial\tau_j\partial\rho_i}(\frac{\delta\tau_j}{\delta\rho_i}+k^2) \right]\delta\rho_j\;.
\ee 
At zero temperature we have $\delta\tau_j/\delta\rho_j=k_{Fj}^2$, 
and taking the value at Fermi level we get:
\be
(\delta h_i^b)_F&=&\sum_j \left[\frac{\partial^2\mH}{\partial\rho_j\partial\rho_i}
+\frac{\partial^2\mH}{\partial\tau_j\partial\rho_i}(k_{Fj}^2+k_{Fi}^2) \right] \delta\rho_j\;.
\ee

The surface term arises from the density-gradient dependence in the Skyrme Hamiltonian density:
\be
\mH^{\nabla}
&=&
C_{nn}^{\nabla}(\nabla\rho_n)^2
+C_{pp}^{\nabla}(\nabla\rho_p)^2
+2C_{np}^{\nabla}\nabla\rho_n\nabla\rho_p)\nonumber\\
&=&
C_{11}^{\nabla}(\nabla\rho)^2+C_{33}^{\nabla}(\nabla\rho_3)^2\;,
\ee
where the coefficients $C^{\nabla}$ are combinations of the usual Skyrme parameters 
(given in Ref.~\cite{camille08}), independent of neutron and proton densities.
A transferred momentum $q$ then induces the nuclear mean-field variation:
\be
\delta h_i^{\nabla}&=&2q^2\sum_j C_{ij}^{\nabla}\delta\rho_j\;.
\ee

Let us finally consider the Coulomb mean-field variation.
In the non-relativistic limit, for nucleons with effective mass $m^*_i$ we have:
\be
\delta (h^c_i)_F&=&\sum_j\left[\frac{4\pi e_ie_j}{q^2} 
\frac{1-xv^*_{Fi}\,\omega/q}{1-\omega^2/q^2}\right]\delta\rho_j\;,
\ee
where $v^*_{Fi}=k_{Fi}^2/m^*_i$; $e_n=0$; $e_p=e={q_e}/\sqrt{4\pi\epsilon_{0}}$.
Only the Coulomb term brings a $x$-dependence in $\delta h_i$.

From the above expressions, we identify:
\be
\mU_{ij}^{(1)}&=&
\left(\frac{\partial^2\mH}{\partial\rho_j\partial\rho_i}
+\frac{\partial^2\mH}{\partial\tau_j\partial\rho_i}(k_{Fj}^2+k_{Fi}^2) \right)\nonumber\\
&&+\left( 2q^2 C_{ij}^{\nabla} \right) \nonumber\\
&&+\left( \frac{4\pi e_ie_j}{q^2} \frac{1}{1-\omega^2/q^2} \right)\\
\mU_{ij}^{(2)}&=&-\frac{4\pi e_ie_j}{q^2} \frac{v^*_{Fi}\,\omega/q}{1-\omega^2/q^2}\;.
\ee


\subsection{\protect\smallskip Vlasov approach with relativistic models}

For relativistic models the one-body Hamiltonian is written in terms of the meson fields
$$h_{i}= \sqrt{(\kv-{\boldsymbol{\cal V}_i})^2 +{m_i^*} ^2} +{\cal V}_{0i},\, i=p,n,$$
where $m_i^*=m-\Gamma_s\phi_0-\tau_i\Gamma_\delta \delta_3$ 
denotes the effective  mass of nucleon $i$ and
$$
{\cal V}_{0i}= \Gamma_v V_0  + \frac{\Gamma_\rho}{2}\, \tau_i b_0
+e A_0 \frac{1+\tau_{i}}{2} + \Sigma_{0}^R\;,
$$
$${\boldsymbol{{\cal V}}}_{i}= \Gamma_v  {\boldsymbol V} +
\frac{\Gamma_\rho}{2}\, \tau_i {\boldsymbol b} + e {\boldsymbol A}
\frac{1+\tau_{i}}{2},
$$ 
with $\tau_i=1\,
(-1)$ for  protons  (neutrons). The
contribution of the  rearrangement term,  due
to the density dependence of the coupling parameters $\Gamma_i$, is given by
$$\Sigma ^R_0=
\frac{\partial\Gamma_v}{\partial \rho}  \rho V_0+
\frac{\partial\Gamma_\rho}{\partial \rho}   \rho_3 \frac{b_{0}}{2}-
\frac{\partial\Gamma_s}{\partial \rho}  \rho_s \phi_0-
\frac{\partial\Gamma_\delta}{\partial \rho}  \rho_{s3} \delta_3\;.$$

The variations of the one-body Hamiltonian which enter the linearised Vlasov equations are:
\begin{equation}
\delta h_i = \delta(m^*_i-m) \frac{m_i^*}{\varepsilon_{i0}} +
\delta{\cal V}_{0i} -\frac{{\bf p}
  \cdot \delta \boldsymbol{\cal V}_i}{\varepsilon_{i0}}\;,
\label{dhi}
\end{equation}
with
$$h_{0 i}\,=\,\sqrt{k^2+{m_i^*}^2}
+ {\cal V}_{0 i}^{(0)}\,=\,\varepsilon_{i0} + {\cal V}_{0 i}^{(0)}$$
and $\delta(m^*_i-m)=-(\Gamma_s\, \delta\phi+\delta\Gamma_s\, \phi_0
+\tau_i\, \Gamma_\delta\,\delta\delta_3+\tau_i\, \delta_3\, \delta\Gamma_\delta)$.
The $x$-dependence of $\delta h_i$ is present through the contribution of the
spatial components of the vector fields: the $\omega$ and $\rho$ mesons and
the electromagnetic field.

Using the linearised equations of the fields, we express the field variations 
in terms of the proton and neutron particle densities and  scalar densities~\cite{stable-modes05} 
and reduce $\delta h_i$ to an expression similar to~(\ref{EQ:dhi-drhoj}).
The coefficients $F^{ij}$ have been defined in~\cite{umodes06,umodes08}.
The nuclear-energy dependence on the transfered momentum
involves the different meson masses,
and is more complex than the Skyrme quadratic expression.
This point will be discussed in more details in Sec.~\ref{SUBSEC:dyn-spino}.

\section{\protect\smallskip Nuclear Matter Properties}\label{SEC:nuclear-matter}

In the present section we will compare the nuclear matter properties 
predicted by all the  models under study. 
We shall consider first the isoscalar properties of the EOS, then the isovector ones.
The spinodal instabilities will also be analysed:
we will address the thermodynamic instability region and direction of phase separation,
and finally  the clusterization properties within the Vlasov approach.


\subsection{\protect\smallskip Symmetric Nuclear Matter }

\begin{figure}[thb]
\begin{center}
\begin{tabular}{cc}
\includegraphics[width =1\linewidth,angle=0]{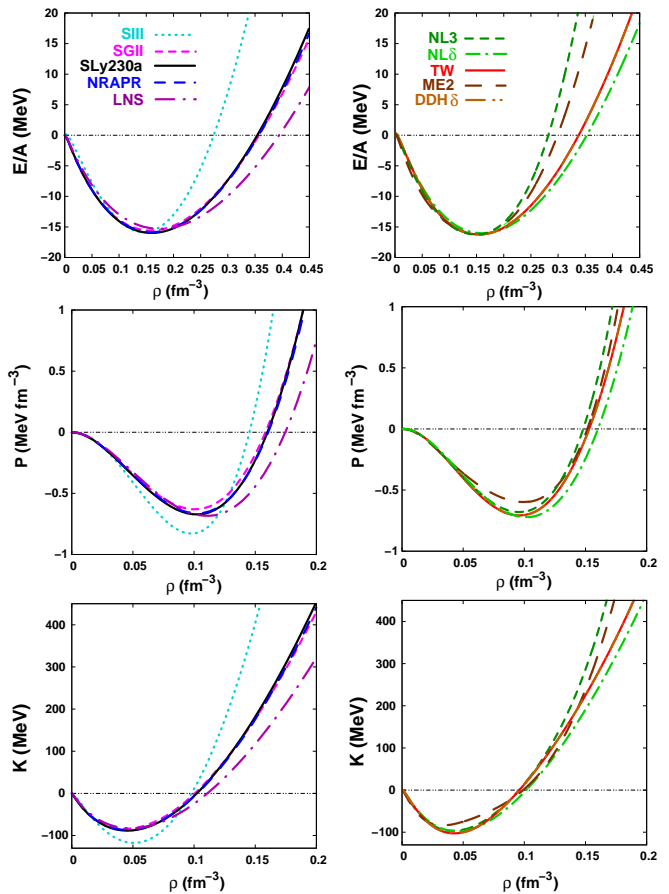}
\end{tabular}
\end{center}
\caption{(Color online) Bulk isoscalar properties of nuclear matter, as a function of the
  baryonic density for non-relativistic (left) and relativistic  (right)
models. From top to bottom: energy per nucleon, pressure and incompressibility.}
\label{FIG:isoscalar-SNM}
\end{figure}

In Fig.~\ref{FIG:isoscalar-SNM} we show some bulk isoscalar properties of nuclear matter
as a function of the baryonic density, namely 
a) the  energy per nucleon,
b) the pressure
$P=\rho^2\partial (E/A)/\partial \rho$, and 
c) the  incompressibility $K=9\partial P/\partial \rho$.
Curves are shown for Skyrme interactions (left) and relativistic models (right). 
Skyrme interactions show similar behaviours between them, except for SIII (stiffer) and LNS (softer).

Globally, the relativistic models present slightly higher binding energies, 
lower saturation densities, 
higher incompressibilities (disregarding SIII) 
and lower effective masses;
however, we can note that NL$\delta$ 
presents isoscalar saturation properties quite similar to the Skyrme ones.
Let us remind that the effective mass has a different meaning in each framework,
as was already stressed in other works~\cite{fuchs06,bcmp07}:
in relativistic models, it includes the contribution of the nucleon scalar self-energy, 
while for the Skyrme interactions it reflects the momentum dependence of the single-particle energy.


\subsection{\protect\smallskip Asymmetric Nuclear Matter }

\begin{figure}[htb]
\begin{center}
\begin{tabular}{cc}
\includegraphics[width=1.\linewidth,angle=0]{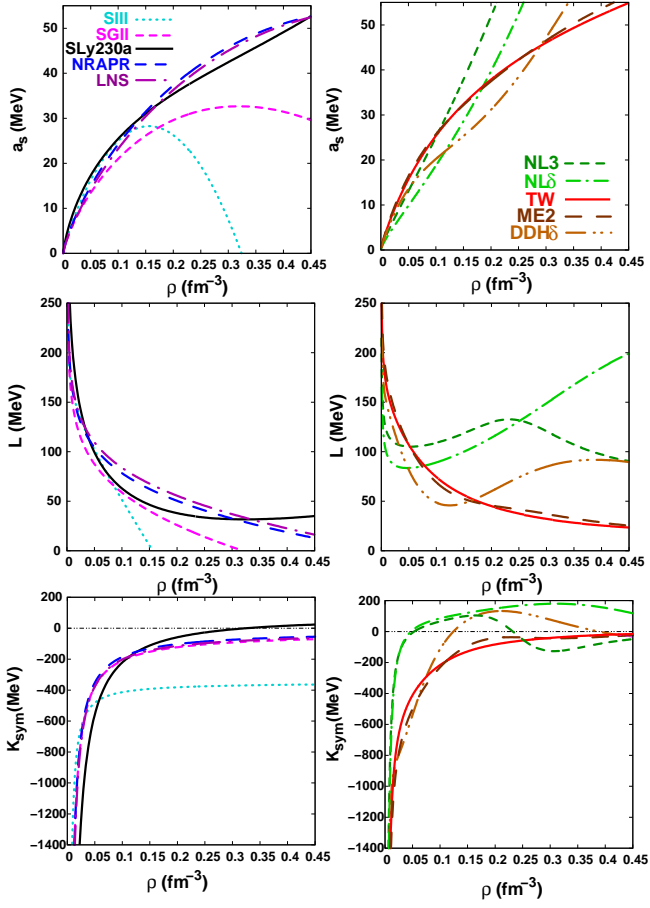}
\end{tabular}
\end{center}
\caption{
(Color online)
Bulk isovector properties of nuclear matter, as a function of the baryonic
density for non-relativistic (left) and relativistic (right) models.
From top to bottom: the symmetry energy and its derivatives with respect to
the density, namely the slope parameter $L$ and the symmetry incompressibility $K_{\rm{sym}}$.}
\label{FIG:isovector-SNM}
\end{figure}
\begin{figure*}[thb]
\begin{center}
\begin{tabular}{cc}
\includegraphics[width=0.5\linewidth,angle=0]{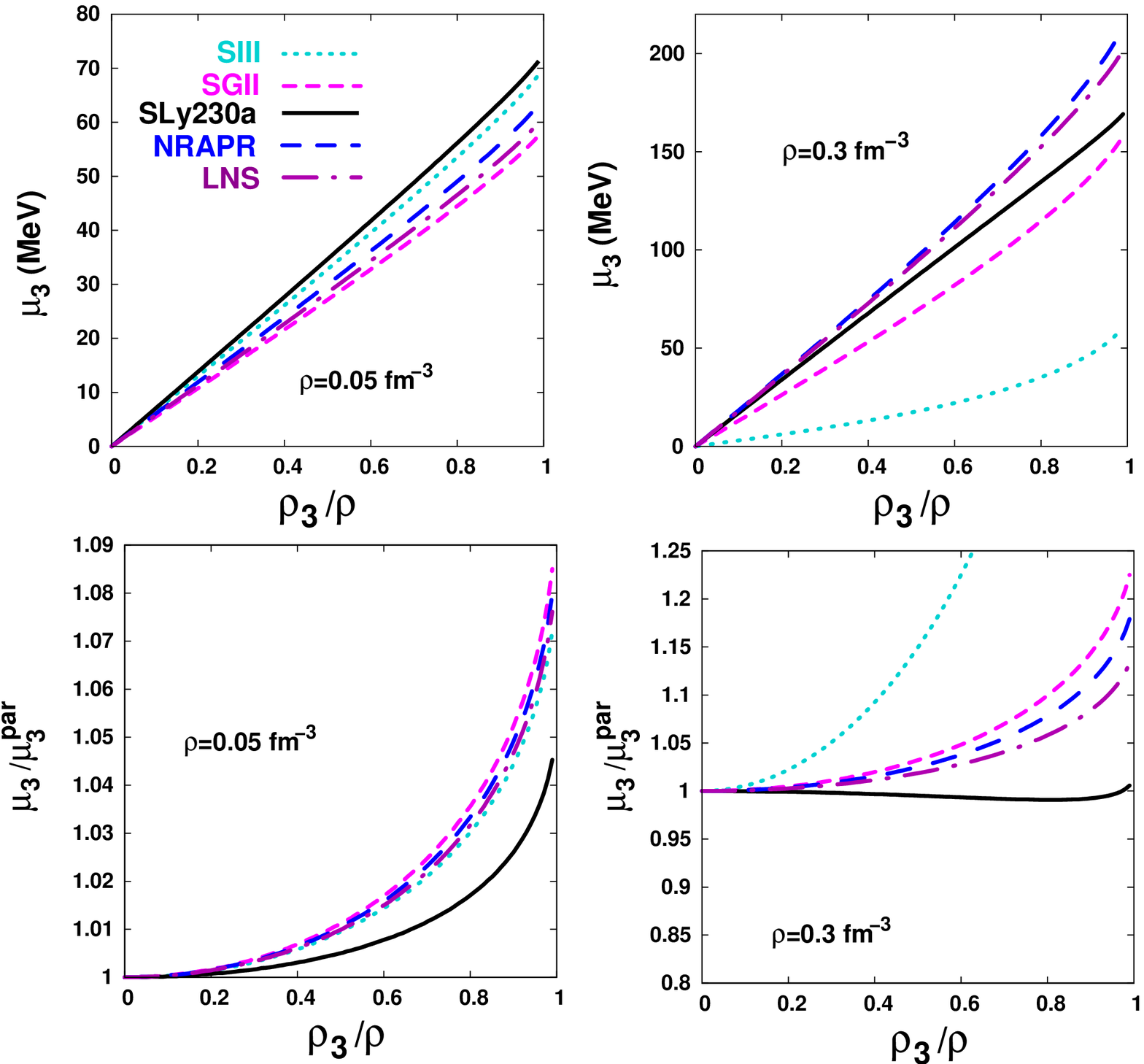}&
\includegraphics[width=0.5\linewidth,angle=0]{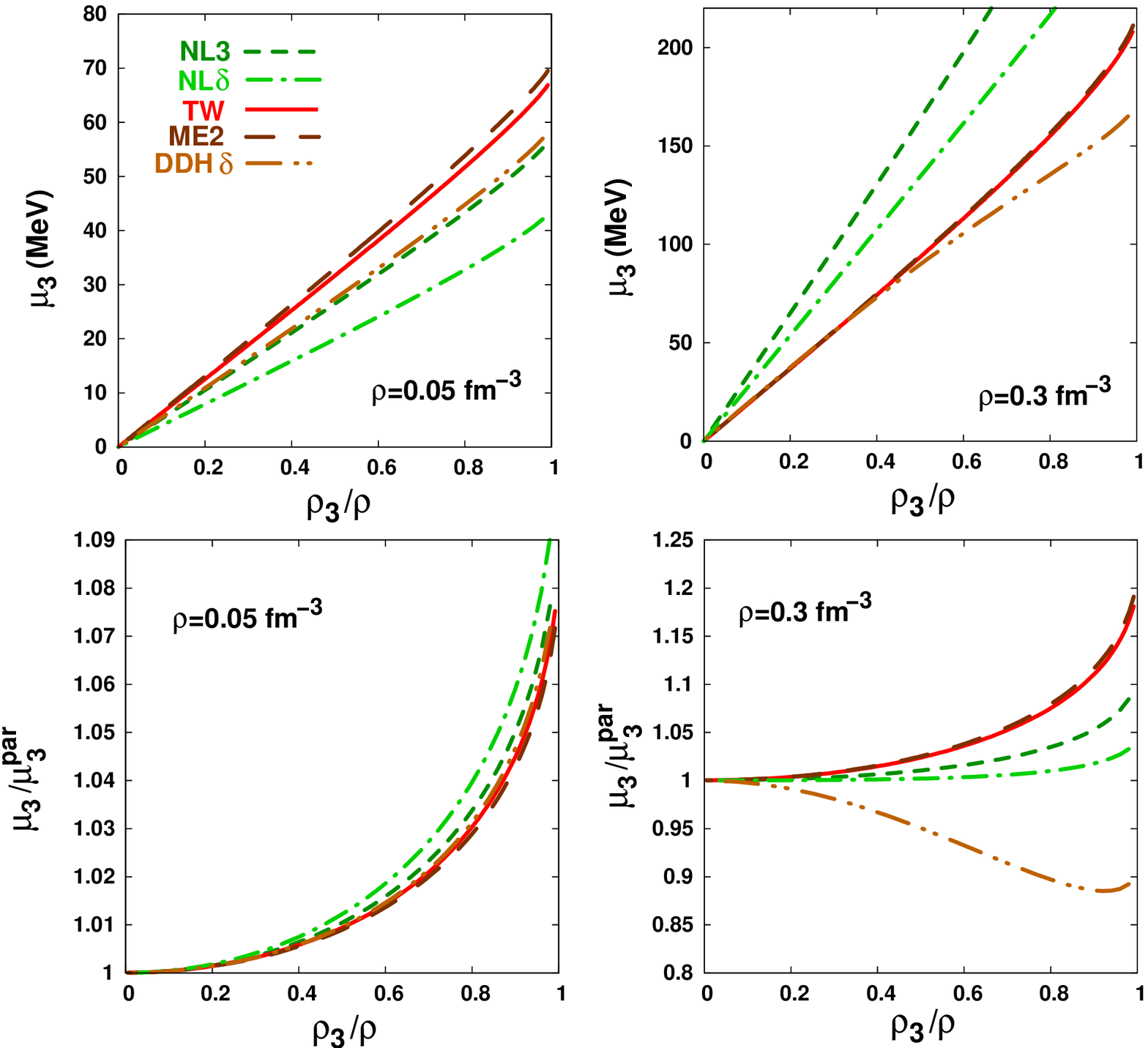}
\end{tabular}
\end{center}
\caption{
(Color online)
Isovector chemical potential $\mu_3=\mu_n-\mu_p$ 
as a function of the asymmetry $y=\rho_3/\rho$
for two values of the total density ($\rho=0.05$ and $0.3$~fm$^{-3}$).
Top: exact value of $\mu_3$. Bottom: the ratio $\mu_3/\mu_3^{\rm{para}}$, where $\mu_3^{\rm{para}}\simeq 4\,a_{\s}\,\rho_3/\rho$.
}
\label{FIG:M3M3para}
\end{figure*}
We will now discuss properties of asymmetric nuclear matter (ANM).
In Fig.~\ref{FIG:isovector-SNM}  we show some properties 
related to the isovector channel of the respective models. 
As expected, larger differences are observed in this channel;
not only between relativistic and Skyrme models, 
but also between different parametrizations inside each framework.
We plot the symmetry energy 

$$a_{\s}=\frac{1}{2}\,\frac{\partial^2 {(E/A)}}{\partial y^2}=\frac{1}{2\rho}\,\frac{\partial^2 {(E/V)}}{\partial y^2},$$
with $y=(\rho_n-\rho_p)/(\rho_n+\rho_p)=\rho_3/\rho$.
Note that this definition can be generalised at finite temperature:
denoting $\cal F$ the free-energy density, it becomes
$$a_{\s}=\frac{1}{2\rho}\,\frac{\partial^2 {\cal F}}{\partial y^2}\;.$$
Figure~\ref{FIG:isovector-SNM} also represents quantities 
related to the first and second density derivatives of the symmetry energy
(respectively denoted by
$a_{\s}^{\prime}=\partial a_{\s}/\partial\rho$ and $a_{\s}^{\pprime}=\partial^2 a_{\s}/\partial\rho^2$), 
according to expressions of common use in the literature~\cite{Baran-PhysRep410}:
the slope parameter
$$L=3\rho_0\, a^{\prime}_{\s}$$
related to the symmetry pressure at saturation,
and symmetry incompressibility
$$K_{\rm{sym}}=9\rho_0^2\, a^{\pprime}_{\s}\;.$$

Among the Skyrme forces, 
the modern parametrizations (SLy230a, NRAPR and LNS) 
show similar values of the symmetry energy in the presented density range. 
As expected, the older SIII parametrization presents atypic features;
it even predicts an isospin instability at  $\rho=0.325$~fm$^{-3}$, as it can
be seen in Fig. \ref{FIG:isovector-SNM} top.
SGII follows an intermediate behaviour. 
It is interesting to see that  NRAPR and LNS, despite different symmetric-matter EOS,
almost coincide in the isovector channel.
With $L$ values at saturation of the order of $60$~MeV, 
these two parametrizations get close to the $L$ range 
estimated from the most recent experimental constraints
(isospin diffusion and isoscaling data): 
$L=88\pm 25$~MeV~\cite{Chen-PRC72, Piekarewicz-PRL95, Shetty-PRC75}, 
while the other three Skyrme paraterizations have too small symmetry-energy slopes at saturation.
The parametrization SLy230a differs from NRAPR and LNS 
by its sharp increase of the symmetry energy at suprasaturation densities.
In the following we will see how these quantities influence
the predictions of the different models for neutron rich matter.

Among the relativistic parametrizations used which do not include the $\delta$-meson, 
the larger differences occur between NL3 
and the models with density dependent couplings: 
TW and DD-ME2, which have very similar behaviours in the isovector channel.
In particular, NL3 has a very hard symmetry energy which increases almost linearly with the density. 
The inclusion of the $\delta$-meson significantly reduces 
the symmetry energy at $\rho<\rho_0$,
but it is also associated with a sharp increase of $a_{\s}$ at higher densities.
Considering the $L$ value at saturation, 
we see a clear separation (roughly a factor of 2) between RMF and DDH models, 
situated on each border of the interval of experimental constraints cited above.

It is interesting to see that TW and DD-ME2 (DDH) behave like NRAPR and LNS (Skyrme) 
for all isovector properties.
However, all other parametrizations show large differences
affecting the three quantities $a_{\s}$, $L$ and $K_{\rm{sym}}$.
The general trend is that relativistic models have a stiffer symmetry energy,
as well as a larger symmetry incompressibility.
For the presented results, several relativistic models 
have a region of positive $K_{\rm{\rm{sym}}}$,
while SLy230a is the only Skyrme parametrization to present such feature.
In the following, we will try to investigate to what extent 
the $\beta$-equilibrium and clusterization properties are affected by these differences.

Let us finally investigate the validity of the parabolic approximation of the isovector EOS,
which is model dependent.
In this approximation, we have a direct link between the symmetry energy
and the isovector chemical potential $\mu_3=\mu_n-\mu_p$,
which determines the matter composition at $\beta$-equilibrium.
Indeed, the parabolic expression of the free-energy density is:
\be
{\cal F} \simeq {\cal F}^{\rm{para}}
&=&{\cal F}_{\s} + \rho\, a_{\s} y^2 
\label{EQ:mu3-as}
\ee
where ${\cal F}_{\s}={\cal F}(\rho,0)$ is the free energy density of symmetric matter.
The corresponding isovector chemical potential is then proportional to $y$, as:
\be
\mu_3^{\rm{para}}
&=&2\frac{\partial {\cal F}^{\rm{para}}}{\partial \rho_3}
= 4 a_{\s} y
\;.
\ee
The parabolic approximation is exact in the limit of small asymmetry,
and actually gives very good predictions for $\mF$ until $y=1$.
However, more significant differences may be obtained 
for the $\rho_3$-derivative leading to $\mu_3$. 
This behaviour is checked on in Fig.~\ref{FIG:M3M3para},
representing both the exact value $\mu_3=2\partial{\cal F}/\partial\rho_3$
and the ratio $\mu_3/\mu_3^{\rm{para}}$.
For this, we have fixed two values of the baryonic density,
$\rho=0.05$ and $0.3$~fm$^{-3}$.
At low densities we mostly confirm the validity of the parabolic approach;
more significant differences are observed at high density.
The dominant trend is to have $\mu_3>\mu_3^{\rm{para}}$, 
due to the kinetic contribution to the symmetry energy;
only the SLy230a and DDH$\delta$ show the opposite behaviour, at high density.


\subsection{\protect\smallskip Thermodynamical spinodal instability}

The liquid-gas phase transition is a well-known feature of the nuclear-matter EOS.
It corresponds to the presence of an abnormal (negative) curvature of the free-energy density $\cal F$
as a function of ($\rho_n,\rho_p$), or equivalently ($\rho,\rho_3$).
The thermodynamical spinodal instability corresponds to the region where the homogeneous matter
is locally unstable against the separation in two infinite homogeneous phases, 
meaning that the surface ${\cal F}(\rho,\rho_3)$ presents a local negative curvature.
This bulk property of nuclear matter is at the origin of the dynamic
instabilities leading to matter clusterization.
We will  consider next the thermodynamical spinodal properties.

\subsubsection{\protect\smallskip Thermodynamic spinodal region}

The spinodal contour is defined by the cancellation of the determinant of the free-energy curvature matrix:
\be
\label{EQ:C-matrix}
C&=&\left(\begin{array}{cc}\mF_{11} & \mF_{13} \\\mF_{31} & \mF_{33}\end{array}\right)\\
\mF_{ij}&=&\frac{\partial^2 \mF}{\partial \rho_i\partial\rho_j}
\ee
where $\rho_1=\rho$.
Inside the spinodal region, the lower eigen-value $C_<$ of this matrix is negative.

\begin{figure}[b]
\begin{center}
\begin{tabular}{cc}
\includegraphics[width =1\linewidth,angle=0]{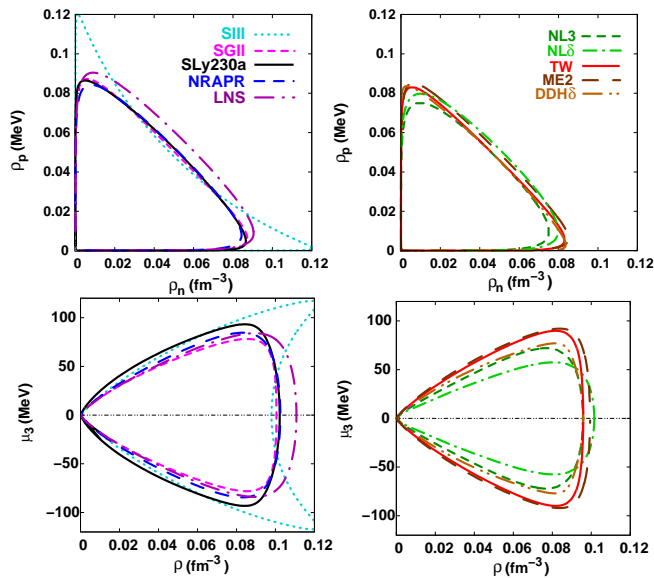}
\end{tabular}
\end{center}
\caption{
(Color online)
Thermodynamic spinodal of infinite nuclear matter for non-relativistic (left)
and relativistic (right) models. The top spinodals are represented on the
$(\rho_p,\rho_n)$ plane and the bottom ones on the $(\mu_3,\rho)$ plane.
}
\label{FIG:spino-thermo}
\end{figure}

The different spinodal contours are shown on Fig.~\ref{FIG:spino-thermo},
using two different representations: the density plane $(\rho_n,\rho_p)$
and the mixed plane $(\rho,\mu_3)$, where isovector differences appear more clearly. 
Although the overall features are similar, 
the trend is that relativistic models predict a smaller instability region,
both in isoscalar and isovector directions.
The isoscalar extension of the spinodal is measured by the density $\rho_{\s}$, 
corresponding to the upper spinodal border for symmetric matter.
The different $\rho_{\s}$ values are reported in  Table~\ref{TAB:Ctilde-as}:
we can verify that they are correlated with the $\rho_0$ values.
As for the isovector behavior of the spinodal contour,
we see that it reaches very high asymmetries with all models.
We can however compare the different extensions obtained in the $\mu_3$ direction.
They are found to reflect the subsaturation-density behavior of the symmetry energy:
indeed, as discussed above, in this density range the isovector chemical potential
can be well-approximated by $\mu_3\simeq 4a_{\s}y$.
For instance, NL$\delta$ has the smallest $a_{\s}$ values at low density,
and therefore presents the narrowest spinodal contour in the $\mu_3$ direction.
We remind that the inclusion of the $\delta$-meson leads to a reduction of $a_{\s}$,
which is observed both with NL$\delta$ and DDH$\delta$.
The $\mu_3$-extension of the spinodal contour is a feature especially relevant 
in the astrophysical context, for the comparison between 
the instability region and the constraint of $\beta$-equilibrium:
this point will be addressed in more details in section~\ref{SEC:stellar-matter}.
\begin{table}[t]
\caption
{
Characterization of the spinodal shape by the contour concavity $\tilde{C}_{\s}$,
depending on the symmetry energy and its density derivatives 
taken at the upper spinodal border of symmetric matter ($\rho=\rho_{\s}$).
}
\begin{center}
\begin{tabular}{lcccccc}
\hline
& $\rho_{0}$ &  $\rho_{\s}$  & $a_{\s}$ & $\rho_{\s} a_{\s}^{\prime}$ & $\rho_{\s}^2 a_{\s}^{\pprime}$  & $\tilde{C}_{\s}$  \\
&(fm$^{-3}$)& (fm$^{-3}$)&(MeV)&(MeV)& (MeV)  &(MeV fm$^3$)\\
\hline
SIII	& 0.145	 & 0.098   & 24.73 & 12.06 &-20.96 &-175.72\\
SGII  	& 0.159	 & 0.100   & 21.06 & 12.41 & -9.33 & 17.15\\
SLy230a	     & 0.160    & 0.102   & 25.73 & 13.21 & -9.70 & 61.60\\
NRAPR& 	 0.161   & 0.103   & 24.60 & 16.50 & -8.24 & 51.22\\
LNS& 	 0.175	 & 0.111   & 24.89 & 16.92 & -8.28  & 46.27\\
\hline
NL3&	 0.148   & 0.096 & 24.04 & 23.40 & 2.86 & 85.98\\
NL$\delta$&0.160 & 0.102 & 19.06 & 18.93 & 3.87      & 81.33\\
TW        &0.153 & 0.096& 24.71 & 15.95 & -9.91 & 29.10\\
DDME2     & 0.152& 0.099& 25.57 & 15.00 & -10.92   & 29.98\\
DDH$\delta$&0.153& 0.096   & 19.80 & 10.38 & -6.03 & 80.25\\
\hline
\end{tabular}
\end{center}
\label{TAB:Ctilde-as}
\end{table}%

Let us now consider the shape of the spinodal contour:
the differences we observe can be caracterized by the convexity of the upper border.
Therefore, we introduced the quantity $\tilde{C}_{\s}$ (hereafter called \emph{contour concavity}),
defined as the convexity of the spinodal contour at point $(\rho=\rho_{\s}, y=0)$~\cite{camille08}:
\be
\tilde{C}_{\s}
&=&\frac{\partial^2C_<(\rho_{\s},0)}{\partial y^2}\nonumber\\
\label{EQ:Ctilde-as}
&=&\frac{2}{\rho_{\s}}
\left[\rho_{\s}^2a_{\s}^{\pprime}+2\rho_{\s} a_{\s}^{\prime}(1-\rho_{\s} a_{\s}^{\prime}/a_{\s})\right]
\;.
\ee
If $\tilde{C}_{\s}$ is positive (negative), 
for a small asymmetry $y$ the point $(\rho_s,y)$ is outside (inside) the spinodal,
meaning a concave (convex) contour.
Equation~(\ref{EQ:Ctilde-as}) gives the relation between $\tilde{C}_{\s}$ 
and the density behavior of the symmetry energy,
involving $a_{\s}$, $a_{\s}^{\prime}$ and $a_{\s}^{\pprime}(\rho)$. 
None of the terms constituting this expression dominates,
as can be appreciated in Table~\ref{TAB:Ctilde-as}:
the symmetry energy, but also its first and second derivatives 
come into play to determine the contour concavity.
Positive $\tilde{C}_{\s}$ values are obtained with all the present models except SIII, 
whose convex shape is due to the large negative values of $a_{\s}^{\pprime}(\rho_s)$.
Concerning the relativistic models,
the models with constant couplings (NL3 and NL$\delta$) 
have larger $a_{\s}^{\prime}$ and positive $a_{\s}^{\pprime}(\rho)$,
giving rise to larger $\tilde C_{\s}$ values.
We can note that the contour concavity is a relevant property 
for the study of non-homogeneities in star matter,
since it determines the sensitivity of the upper spinodal border to the specific composition
that will be imposed by the $\beta$-equilibrium.

\subsubsection{\protect\smallskip Thermodynamic instability direction}
\begin{figure}[t]
\begin{center}
\includegraphics[width=1\linewidth,angle=0]{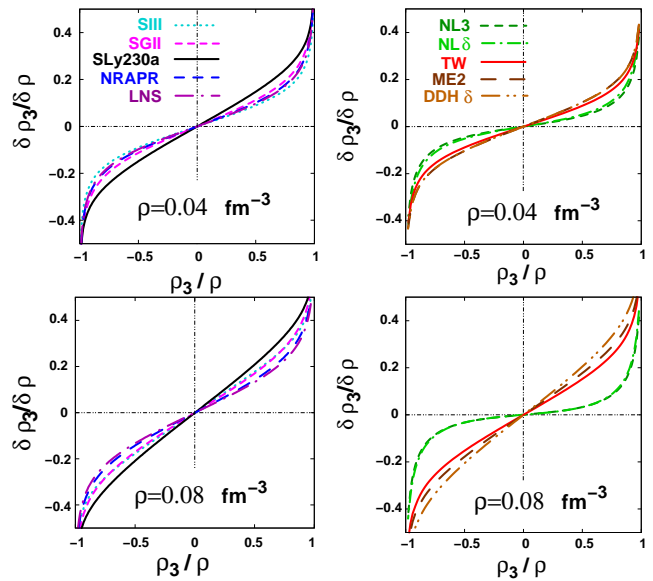}\\
\end{center}
\caption{
(Color online)
Thermodynamic instability direction as a function of the asymmetry, for two fixed total densities.
Results are given for Skyrme (left) and relativistic (right) models.
}
\label{FIG:dr3sdr-thermo1}
\end{figure}

The thermodynamic instability direction is 
the direction of minimal free-energy curvature,
given by the eigen-vector of matrix~(\ref{EQ:C-matrix}) associated with $C_<$.
It is related to the phenomenon of isospin distillation,
which usually leads to the formation of a dense phase more symmetric than the dilute one.
We express this direction as the ratio $\delta\rho_3/\delta\rho$, 
giving the deviation with respect to the isoscalar direction.
The eigen-vector $(\delta \rho, \delta \rho_3)_<$ satisfies:
\be
\frac{\delta \rho_3}{\delta\rho}&=&\frac{C_<-{\cal F}_{11}}{{\cal F}_{13}}
\;.
\ee
This ratio is zero in the case of symmetric matter, where the instability direction is purely isoscalar.
For extremal asymmetry $y=\pm 1$, it obeys the limit conditions $\delta\rho_3/\delta\rho=\pm1$,
which constrains the behavior of $\delta\rho_3/\delta\rho(y)$ at high asymmetry.
However, for moderate values of asymmetry, 
the evolution of the instability direction is nearly linear with respect to $y$,
as illustrated in Fig.~\ref{FIG:isovector-SNM}.
In this region, the isospin-distillation properties of the different effective forces
can be characterized by a number $\tilde\delta$ such that
\be
 \frac{\delta\rho_3}{\delta \rho}&=& \tilde\delta\, y+{\cal O}(y^3)\\
\tilde\delta&=&\frac{a_{\s}-\rho a_{\s}^{\prime}}{a_{\s}-\rho{\cal F}^{\pprime}_{\s}/2}
\;.
\ee

\begin{figure}[t]
\begin{center}
\includegraphics[width=1\linewidth,angle=0]{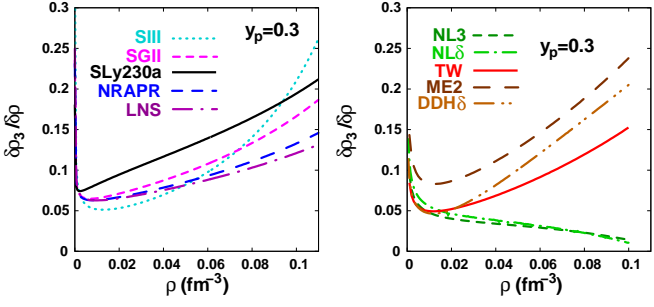}\\
\end{center}
\caption{
Thermodynamic instability direction as a function of the density, 
at proton fraction $Y_p=0.3$, for non-relativistic (left) and relativistic (right) models.
}
\label{FIG:dr3sdr-thermo}
\end{figure}

On Fig.~\ref{FIG:dr3sdr-thermo}, we show the density evolution of $\delta\rho_3/\delta\rho$ 
at a fixed proton fraction $Y_p=0.3$, corresponding to the asymmetry $y=\rho_3/\rho=0.4$.
Note that this reflects the $\tilde{d}$ values according to the good approximation 
$\tilde{d}\simeq (\delta\rho_3/\delta\rho)/y$.
For all models, $\delta\rho_3/\delta\rho$ is lower than $y=0.4$:
this is the normal distillation effect.
It is seen that the relativistic models with constant coupling, 
independently of containing or not the $\delta$-meson, predict a much larger distillation effect 
(smaller ratio  $\delta\rho_3/\delta\rho$):
this is due to the quasi-linear behavior of $a_{\s}(\rho)$ (see Fig.~\ref{FIG:dr3sdr-thermo1}),
which leads to low values of $\tilde{d}$.
In contrast, the DDH models behave like the Skyrme forces:
both show a reduction of the distillation effect with density, 
while NL3 and NL$\delta$ present the opposite behavior 
(which was also noticed in~\cite{inst06,cp07}).


\subsection{\protect\smallskip Dynamical spinodal instabilities }\label{SUBSEC:dyn-spino}

The bulk liquid-gas instability properties we have discussed 
do not manifest themselves directly
in nuclear multifragmentation and compact-star matter,
due to the role of the Coulomb interaction and the surface tension.
However, they induce instabilities against finite-size density fluctuations,
leading to the decomposition of the homogeneous matter into a clusterized medium~\cite{Pethick-NPA584}.
We now use the formalism presented in section~\ref{SEC:vlasov} 
to study the Vlasov unstable modes, 
considering plane-wave density fluctuations of wave-number $\qv$.
The dispersion relation is defined by Eq.~(\ref{eq:rd}), with $\omega=i/\tau$; 
$\tau$ is the time constant which characterizes the initial growth
of the density fluctuation.

\begin{figure}[b]
\begin{center}
\includegraphics[width =1\linewidth,angle=0]{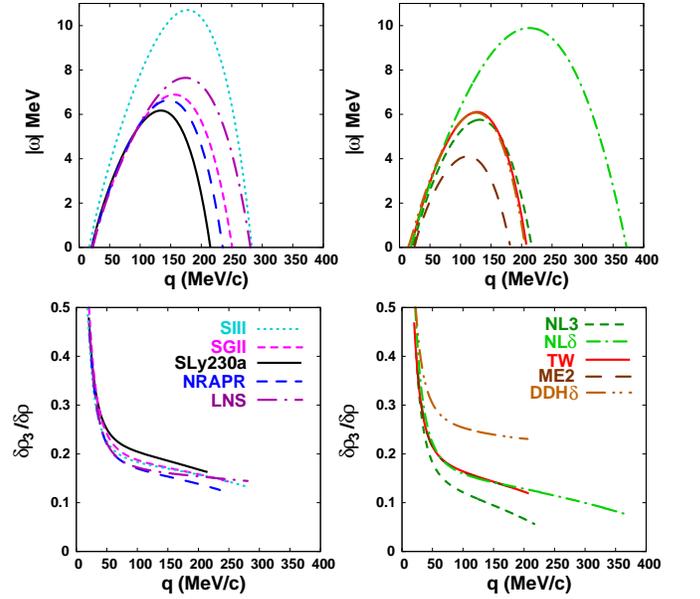}
\end{center}
\caption{
(Color online)
Finite-size instabilities:
dispersion relation for the unstable modes
at $\rho=0.05$~fm$^{-3}$, $Y_p=0.3$, $T=0$,
for Skyrme (left) and relativistic models (right).
Top: modulus of the imaginary frequency $|\omega|$ as a function of the wave number.
Bottom: direction of the unstable mode $\delta\rho_3/\delta\rho$ in the
density plane. The dotted lines show $\rho_3/\rho=0.4$ corresponding to the
proton fraction considered.  }
\label{FIG:dispersion}
\end{figure}

On Fig.~\ref{FIG:dispersion},
we compare the unstable modes obtained within the different models 
for fixed average densities ($\rho=0.05$~fm$^{-3}$, $Y_p=0.3$), 
as a function of the wave number.
The top figures represent the growth rate $|\omega|=1/\tau$, 
and the bottom ones the  direction of the mode in the density plane given by $\delta\rho_3/\delta\rho$. 

Considering the top part of Fig.~\ref{FIG:dispersion},
we first see that relativistic models are usually characterized by a reduced instability.
To the noticeable exception of NL$\delta$,
both the growth rate and the upper border of the unstable $q$ interval
are smaller within the relativistic models.
The bottom part of Fig.~\ref{FIG:dispersion},
gives the phase-separation direction ${\delta \rho_3}/{\delta\rho}$
associated with the dynamical modes.
All curves decrease with $q$, due to the Coulomb effect:
at low $q$, the strong Coulomb contribution quenches the proton-density fluctuation,
imposing large values of ${\delta \rho_3}/{\delta\rho}$.
This leads to the so called {\em anti-distillation effect}, 
corresponding to ${\delta \rho_3}/{\delta\rho}>\rho_3/\rho$,
namely a dense phase more neutron-rich than the homogeneous matter:
it is obtained here below $q\sim30$~MeV/c.
For higher $q$ values, we recover the normal distillation effect with all the models.
Comparing the present dynamic results with the bulk instability direction
(Fig.~\ref{FIG:dr3sdr-thermo}), 
we see that the behavior of the relativistic models deserves a comment.
Although NL3 remains the model with the strongest distillation effects,
the hierarchy of the other curves is widely rearranged when dynamical instabilities are considered:
the weaker distillation effect is now obtained with models including the $\delta$ meson,
both NL$\delta$ and DDH$\delta$. 

\begin{figure}[t]
\begin{tabular}{ccc}
\includegraphics[width=1\linewidth]{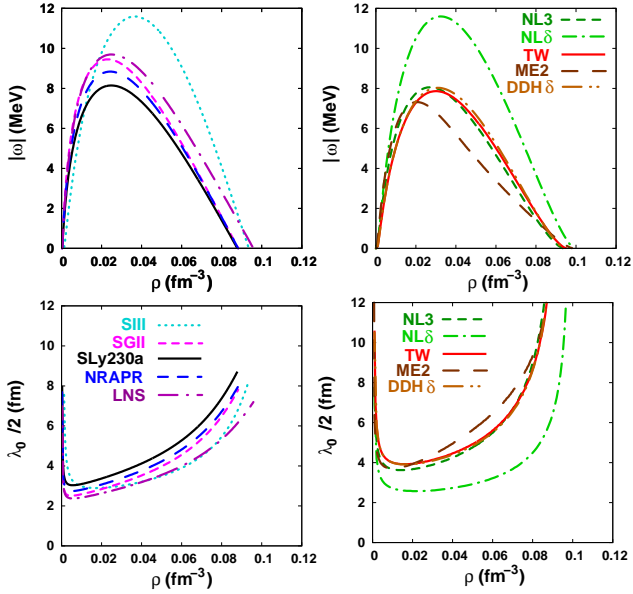}
\end{tabular}
\caption{Most unstable modes for np matter with proton fraction $Y_p=0.3$,
for Skyrme parametrizations (left) and  relativistic models (right).
Top: growth rates. Bottom: associated cluster size.}
\label{most_inst}
\end{figure}

Let us now study the most unstable mode, namely the mode of largest growth rate,
which drives the system to the non-homogeneous phase. 
The associated wave-number characterizes the size of the primary clusters 
formed in spinodal decomposition,
which we can define as the half wavelength of the fastest amplified mode.
In Fig.~\ref{most_inst} we show the growth rates and associated cluster size of
the most unstable modes. 
We show results for $Y_p= 0.3$, a proton fraction 
close to that of $\beta$-equilibrium matter with neutrino trapping
and to the asymmetry values that could be involved in future multifragmentation experiments
with radioactive beams (for instance $^{132}$Sn has $Y_p=0.379$).
Consistently with the observations of Fig.~\ref{FIG:dispersion},
as a general rule the Skyrme parametrizations predict 
larger growth-rates and smaller clusters than the relativistic models.
The hierarchy between the different parametrizations is also essentially conserved:
among Skyrme forces, SLy230a gives the largest clusters and LNS the smallest ones;
among the relativistic models, NL$\delta$ predicts particularly small sizes
and DD-ME2 gives the largest clusters.
Furthermore, we can notice features appearing with the density evolution.
Firstly, going to lower densities, 
the Skyrme cluster sizes decrease more neatly than the relativistic ones.
Thus, the minimal cluster sizes obtained in each framework are
of $\sim 2.5$~fm (Skyrme) and $\sim 4$~fm (relativistic models).
Secondly, larger sizes are reached near the border of the unstable region
($\sim 8$~fm with Skyrme, and beyond $10$~fm with the relativistic models).
These features accentuate the trend according which 
the relativistic models predict larger clusters.

The different $q$-dependences originate in the finite-range part of the nuclear force,
which was introduced in Sec.~\ref{SEC:vlasov}.
Since the direction of the density fluctuations is essentially isoscalar,
we can characterize the energy cost of the density gradient
by the quantity $C_{11}^{\nabla}q^2$ (for the nuclear contribution).
This quadratic expression is exact for Skyrme models,
but the $q$-dependence is more complex for relativistic models:
performing a Taylor expansion in powers of $f_i=q^2/m_i^2$ ($m_i$ denoting the meson masses),
we obtain a density-dependent $C_{11}^{\nabla}$ coefficient.
The values of $C_{11}^{\nabla}$ are listed in Table~\ref{TAB:fi}:
for relativistic models, they are given at $\rho=0.05$~fm$^{-3}$, 
together with the $f_i$ values.

\begin{table*}[hbt]
  \centering
\caption{
Dependence of the nuclear energy on the transfered momentum
characterized by the $C_{11}^{\nabla}$ coefficients.
These coefficients are constant for Skyrme models.
For relativistic models, they are given at $\rho=0.05$~fm$^{-3}$,
together with the corresponding parameters:
$f_i=g_i^2/m_i^2$ for $i=\sigma,\, \omega,\, \delta$,  
and $f_\rho=g_\rho^2/(4\,m_\rho)^2$.
}
  \begin{tabular}{cc|cccccc}
\hline
    Skyrme  & $C_{11}^{\nabla}$ 	&Relativistic 	& $C_{11}^{\nabla}$ &  $f_s$ &  $f_v$ & $f_\rho$ & $f_\delta$\\
    	          & (MeV.fm$^5$) 		& 			& (MeV.fm$^5$) & (fm$^2$) & (fm$^2$) & (fm$^2$) & (fm$^2$) \\          
\hline
SIII      	& 63.0 &	NL3			& 99.2 & 15.73 & 10.53 & 1.34 & 0 \\
SGII      	& 54.8 &	NL$\delta$	& 43.0 & 10.33 & 5.42 & 3.15 & 2.5 \\
SLy230a  	& 77.7 &	TW 			& 115.9 & 18.97 & 14.64 & 1.79 & 0 \\
NRAPR    	& 64.1 &	DD-ME2		& 107.7 & 18.50 & 13.99 & 1.94 & 0 \\
LNS        	& 43.8 &	DDH$\delta$	& 115.9 & 18.97 & 14.64 & 4.16 & 2.96 \\
\hline
  \end{tabular}
  \label{TAB:fi}
\end{table*}

\begin{figure}[b]
\begin{center}
\includegraphics[width =1\linewidth,angle=0]{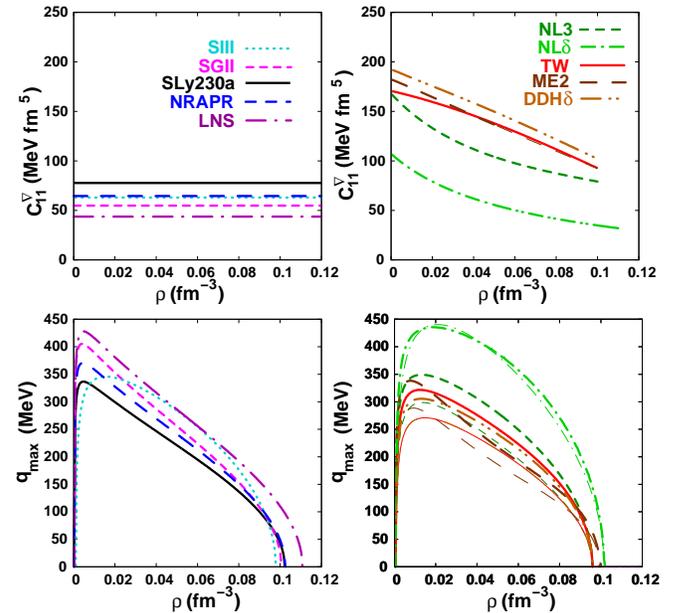}
\end{center}
\caption{
(Color online)
$C_{11}^\nabla$ and $q_{\rm{max}}$ for Skyrme (left) and
relativistic (right) models, for symmetric matter.
For relativistic models we compare $q_{\rm{max,quad}}$ (thick lines)
obtained from the quadratic expansion
with the exact value $q_{\rm{max}}$ (thin lines).}
\label{FIG:qmax}
\end{figure}

The correspondence between $C_{11}^{\nabla}$ 
and the maximal unstable momentum $q_{\rm{max}}$ is shown on Fig.\ref{FIG:qmax} 
for symmetric matter, as a function of the density.
Neglecting the Coulomb interaction (whose contribution is vanishing for the $q$ range of interest),
the quadratic expression of the $q$-dependence leads to the following relation:
\be
q_{\rm{max,quad}}^2 &=& \frac{|C_<|}{2C_{11}^{\nabla}}\;.
\ee
For the relativistic models, both $q_{\rm{max,quad}}$ and the exact $q_{\rm{max}}$ values 
are shown on the figure: the exact values are smaller than the values calculated
in the quadratic approximation.
The different $C_{11}^{\nabla}$ values are seen to explain 
the different cluster sizes obtained between parametrizations of a same framework
(although atypic $|C_<|$ values can distort the correspondence between $C_{11}^{\nabla}$
and the favored $q$ values).
However, it is not sufficient to explain the difference between relativistic and Skyrme models:
indeed, in the relativistic case, the larger $C_{11}^{\nabla}$ are compensated by larger $|C_{<}|$,
leading to values of $q_{\rm{max,quad}}$ similar to the Skyrme ones.
The larger cluster sizes predicted by the relativistic models 
are due to the non-quadratic part of their $q$-dependence.


\bigskip
\section{\protect\smallskip Stellar Matter}
\label{SEC:stellar-matter}

In the last section, we have discussed nuclear-matter properties
in the framework of Skyrme and relativistic models.
We now want to investigate the consequences of the different features we have obtained
in the context of compact-star physics.
In the first part, we discuss the EOS of homogeneous matter at $\beta$-equilibrium
(disregarding the liquid-gas instabilities), 
considering the possibility of neutrino trapping.
In the second part, we address the implications of the dynamic instabilities
for compact-star properties: width of neutron-star crusts
and non-homogeneities in the cores of type-II supernovae.


\subsection{\protect\smallskip Homogeneous $\beta$-equilibrium matter }

The $\beta$-equilibrium conditions impose the following relations
between the chemical potentials of the particles:
$$\mu_e-\mu_{\nu_e}=\mu_n-\mu_p=\mu_3\;,$$
where $\mu_{\nu_e}=0$ for neutrino-free matter.
Muons are present if they can be in chemical equilibrium with the electrons, satisfying
$$\mu_\mu-\mu_{\nu_\mu}=\mu_e-\mu_{\nu_e}\;;$$
the muon onset thus occurs when
$$\mu_e-\mu_{\nu_e}=m_{\mu}\;.$$
For neutrino-free matter, this condition reduces to $\mu_e=\mu_3=m_\mu$,
from which we determine the muon onset density $\rho_{\mu-\rm{onset}}$
for the different nuclear models.
The corresponding values are given in Table~\ref{TAB:muon}.
In most cases, $\rho_{\mu-\rm{onset}}$ belongs to the interval $[0.112;\,0.124]$ fm$^{-3}$;
we can notice the higher values 
obtained from SGII and the relativistic models with $\delta$ meson.
This is a consequence of the symmetry-energy behavior in this range of densities.
A higher symmetry energy increases the proton fraction at $\beta$-equilibrium:
the electron chemical potential is then higher, 
and the muon onset is reached more easily.
We can verify on Fig.~\ref{FIG:isovector-SNM} that, for $\rho \sim 0.11$ fm$^{-3}$, 
the symmetry energy curves have very similar values, 
except precisely for SGII and the models with $\delta$ meson for which $a_{\s}$ is smaller.

\begin{table}[b]
\caption{Baryonic density at muon onset. For SIII the muons disappear at
  densities larger than 0.23 fm$^{-3}$.}
\begin{tabular}{lc}
\hline
model & $\rho_{\mu-\rm{onset}}$ (fm$^{-3}$)\\
\hline
SIII&0.119\\
SGII&0.146\\
SLy230a&0.121\\
NRAPR&0.117\\
LNS&0.124\\
\hline
NL3&0.112\\
NL$\delta$ & 0.142\\
TW&0.115\\
DDME2&0.114\\
DDH$\delta$&0.166 \\
\hline
\end{tabular}
\label{TAB:muon}
\end{table}

\begin{figure}[t]
\begin{center}
\begin{tabular}{cc}
\includegraphics[width=1\linewidth,angle=0]{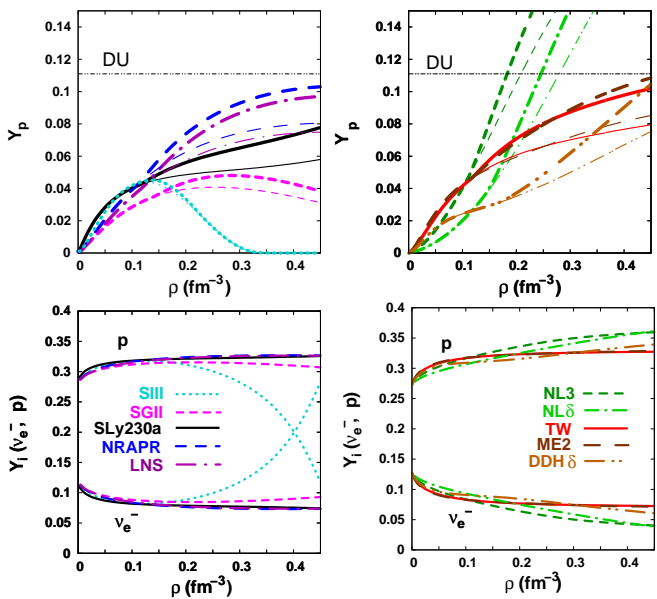}&
\end{tabular}
\end{center}
\caption{
(Color online)
$\beta$-equilibrium for homogeneous matter 
at $T=0$, for non-relativistic (left) and relativistic (right) models.
Top: neutrino-free matter; 
the proton fraction is given considering the muon onset (thick line) 
or considering only electrons (thin line).
Bottom: proton and neutrino fractions for a constant lepton fraction $Y_l=0.4$. 
The direct URCA threshold for matter without muons is indicated in the top figures by a thin line (DU).
}
\label{FIG:beq}
\end{figure}

In Fig.~\ref{FIG:beq} top, 
we plot the proton fractions at $\beta$-equilibrium for neutrino-free matter.
Results are shown taking muons into account  
or considering only electrons. 
As noticed earlier, the proton fraction at $\beta$ equilibrium essentially reflects 
the symmetry-energy $a_{\s}(\rho)$: see Fig.~\ref{FIG:isovector-SNM}.
For most models, the proton fraction increases quite softly with density,
reaching a $Y_p$-range of $\sim[0.08;0.11]$ at $\rho=0.45$~fm$^{-3}$.
Two kinds of atypic behaviors are observed, following the symmetry-energy features.
(1) The two older Skyrme parametrizations show a rise and fall of the proton fraction with density,
eventually leading to pure neutron matter. 
(2) On the opposite, the relativistic models with constant couplings 
show a very sharp increase of the proton fraction with density. 
DDH-$\delta$ also predicts a quite sharp increase of $Y_p$ at high densities,
due to the effect of the $\delta$-meson on $a_{\s}(\rho)$.

However, let us consider the three modern Skyrme forces: SLy230a, NRAPR and LNS.
Despite very close values of the symmetry energy,
the proton fractions obtained with SLy230a are much lower.
Looking at Fig.~\ref{FIG:M3M3para}, we see this is a consequence 
of the different behaviors of the ratio $\mu_3/\mu_3^{\rm{para}}$.
For similar values of the symmetry energy, at high asymmetry
the $\mu_3$ value is lower for SLy230a than for NRAPR and LNS:
the $\beta$-equilibrium is thus realised for lower $Y_p$.
The same effect can be observed among the relativistic models,
comparing DDH-$\delta$ to the other DDH models.

The cooling of neutron stars may occur through a direct/indirect URCA process~\cite{urca}. 
Since the first predicts a too fast cooling, 
models which allow it are not adequate for the description of asymmetric matter. 
If the muon onset is not taken into account, 
the critical proton faction that allows direct URCA is $y_{DU}=1/9$. 
In the presence of muons, this fraction is increased to~\cite{klahn06}
$$y_{DU}=\frac{1}{1+\left(1+x_e^{1/3}\right)^3},\quad x_e=\frac{\rho_e}{\rho_e+\rho_\mu}\;.$$
Only the relativistic models with constant coupling present proton-fractions large enough
for a direct URCA process in the range of density we show:
the sharp $a_{\s}$ evolution of these models
is thus in contradiction with the neutron-star cooling observations.

On the bottom of Fig~\ref{FIG:beq}, 
we consider the $\beta$-equilibrium in matter with trapped neutrinos,
taking a constant  lepton fraction $Y_l=0.4$.
Due to the presence of electronic neutrinos, for the range of density that we consider
the muon fraction at equilibrium is vanishingly small:
in this situation we will include only the constituents $n$, $p$, $e$, $\nu_e$.
On the figure, we show the proton and neutrino fractions.
The most striking feature is that all models (except SIII) give similar predictions,
with a nearly constant proton fraction in the range $\sim[0.3;0.35]$:
when the lepton fraction is fixed, 
the dependence of matter compositions on the symmetry energy becomes very weak. 
 

\subsection{\protect\smallskip Clusterization of stellar matter}

In the present section, 
we discuss the formation of non-homogeneities in compact-star matter,
considering consequences for neutron-star crust and supernova core.
For neutron-star crust, we consider neutrino-free matter at $T=0$;
for the supernova context, we have to include finite temperature,
and it is relevant to consider the effect of neutrino trapping.
Finite temperature results are given only for Skyrme models and NL3.

\begin{figure}[t]
\begin{center}
\begin{tabular}{cc}
\includegraphics[width=1\linewidth,angle=0]{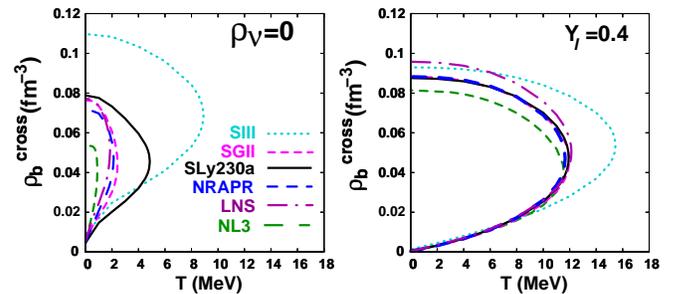}
\end{tabular}
\end{center}
\caption{
(Color online)
Total density at the crossing between the instability region and the $\beta$-equilibrium condition,
as a function of the temperature.
Left: $\beta$-equilibrium in neutrino-free matter. 
Right: $\beta$-equilibrium at constant lepton fraction $Y_l=0.4$.
}
\label{FIG:Trhocros}
\end{figure}


Although the spinodal region almost reaches pure neutron matter at $T=0$,
it is limited to more symmetric matter as the temperature increases,
until it disappears for a limiting value of $T$~\cite{camille08}.
We wish to determine under which conditions compact-star matter at $\beta$-equilibrium
reaches the region of instability against cluster formation.
On Fig.~\ref{FIG:Trhocros}, we plot the total density 
at the crossing between the instability region and the $\beta$-equilibrium condition,
as a function of the temperature. 
We call $T_{\rm{cross}}$ the maximal temperature for which this crossing occurs.
Two cases are considered:
neutrino-free matter and matter with trapped neutrinos ($Y_l=0.4$).
A strong model-dependence is observed in the case of neutrino-free matter,
where the $\beta$ equilibrium involves very neutron-rich matter.
However, two common features can be drawn:
(i) for all models, the instability region is crossed at $T=0$;
(ii) for most models, we obtain $T_{\rm{cross}}<3$~MeV. 
With SIII, higher temperatures are reached because of the larger (irrealistic) instability.
For the other parametrizations,
$T_{\rm{cross}}$ is correlated with the symmetry energy at low density:
the highest value is thus obtained with SLy230a, and the lowest one with NL3.
Indeed, higher $a_{\s}$ values lead to a more symmetric composition,
deeper inside the instability region.
The reciprocal analysis can be made in terms of $\mu_3$:
a spinodal region with a large $\mu_3$ extension is more easily reached at $\beta$ equilibrium
\footnote{
Let us consider a point $S$ of the spinodal contour, 
with a given density and proton fraction determining the electron chemical potential $\mu_e^S$.
The isovector chemical potential in this point is $\mu_3^S$.
For this density, the $\beta$-equilibrium is inside the spinodal region if $\mu_e^S<\mu_3^S$.
A crossing between the $\beta$-equilibrium condition and the instability region 
is then favored if the spinodal contour takes large $\mu_3$ values.
}.
In the case of matter with trapped neutrinos, 
a crossing is obtained  until $T > 10$~MeV for all models:
indeed, the proton fractions are now larger than $0.3$.
The differences we observe in the upper-density crossing
reflect the various $\rho_{\s}$ values 
(upper border of the thermodynamic spinodal for symmetric matter).

\begin{table}[t]
  \centering
  \begin{tabular}[c]{lccccc}
    \hline
   & $Y_{\nu  }=0$ & & &$Y_{L}=0.4$  &\\
 \cline{2-3} \cline{5-6}
model &$\rho_{\rm cross,out}$&$\rho_{\rm cross,in}$&&$\rho_{\rm
  cross,out}$&$\rho_{\rm cross,in}$\\
 & $\times 10^{-2}$(fm$^{-3}$) &(fm$^{-3}$) && $\times 10^{-2}$(fm$^{-3}$) &(fm$^{-3}$) \\
    \hline
SIII&	0.877&	0.110	&  & 0.130&	0.093\\
SGII&	0.418&	0.076& &	0.029&	0.088\\	
SLy230a&0.459&	0.079& &	0.035&	0.088\\	
NRAPR&	0.475&	0.072&  & 0.031&	0.088\\
LNS&	0.543&	0.077& &	0.037&	0.096\\
\hline
    NL3 & 0.553 & 0.053 & & 0.083 & 0.081\\
    NL$\delta$ & 0.442 & 0.057 &  &0.086 & 0.090\\
    TW & 0.915 & 0.075 & & 0.108 & 0.084\\
    DD-ME2 & 0.610 & 0.072 &  &0.060 & 0.083\\
    DDH$\delta$ & 0.776 & 0.079 &  &0.098 & 0.084\\
    \hline
  \end{tabular}
  \caption{
Predicted density at the outer ($\rho_{\rm cross,out}$)  and inner edge ($\rho_{\rm cross,in}$)
of the crust of a compact star at zero temperature,
as defined by the crossing between the dynamical instability region
and the $\beta$-equilibrium condition for homogeneous, neutrino-free stellar matter.}
\label{tab:inner}
\end{table}

In Table \ref{tab:inner} we give the crossing densities at $T=0$,
for the neutrino-free and neutrino-trapping cases.
We denote $\rho_{\rm{cross,in}}$ ($\rho_{\rm{cross,out}}$) the density at the high (low) density crossing point.

The value of $\rho_{\rm{cross,in}}$ obtained for neutrino-free matter at $T=0$
provides the lowest estimation for the density at the inner border of the crust, $\rho_{\rm{crust,in}}$.
Indeed, the finite-size instability region is the minimal region 
where the matter at thermodynamic equilibrium is in a clusterized shape,
so it is contained by the crust. 
We expect $\rho_{\rm{cross,in}}$ to be a good approximation to $\rho_{\rm{crust,in}}$.
Most of the models we present give $\rho_{\rm{cross,in}}\sim 0.075$~fm$^{-3}$.
The lower values obtained with NL3 and NL$\delta$ can be related
to their larger spinodal-contour concavity $\tilde{C}_{\s}$, discussed in Sec.~\ref{SEC:nuclear-matter}.
Nuclear matter in the crust of neutron stars has been studied  recently~\cite{gogelein1,gogelein2} 
both within RMF, DDH and self-consistent Skyrme Hartree-Fock. 
For matter at $\beta$-equilibrium,
the transition densities to the homogeneous phase predicted in these works are: 
$0.085$~fm$^{-3}$ (Skyrme), $0.061$~fm$^{-3}$ (DDH) and  $0.072$~fm$^{-3}$ (RMF).
These values are in reasonable agreement with the numbers given in Table~\ref{tab:inner}, 
although according to our results 
with DDH models the transition density should be higher than $\sim0.07$~fm$^{-3}$.
We can also point out that within each framework the values 
depend on the properties of the chosen parametrization. 

On the other hand, the lower-density crossing point
$\rho_{\rm{cross,out}}$ has little significance for the crust,
since the outer crust border does not correspond to a transition to homogeneous matter.
At zero temperature,
the very low density matter is always made of clusters,
and the concept of homogeneous nuclear matter breaks down.
Note however that both $\rho_{\rm{cross,in}}$ and $\rho_{\rm{cross,out}}$ 
may be of physical interest at finite temperature,
for the formation of non-homogeneities in supernova cores.
The $\rho_{\rm{cross,out}}$ values tend to be larger with the relativistic models,
which is another manifestation of their reduced instability region.

Let us now consider the crossing densities for matter with trapped neutrinos, at $Y_l=0.4$.
They are given at $T=0$ for a direct comparison with the case of neutrino-free matter;
however, we should remark that neutrino-trapping 
occurs in the early stage of neutron-star evolution,
involving finite temperatures.
The value of $\rho_{\rm{cross,in}}$ is higher than for neutrino-free matter, except for SIII:
this also reflects spinodal-contour concavity,
since with trapped neutrinos the $\beta$ equilibrium is shifted towards symmetric matter.
The value of $\rho_{\rm{cross,out}}$ is typically one order of magnitude lower than for neutrino-free matter.

\begin{figure}[t]
\begin{center}
\begin{tabular}{cc}
\includegraphics[width=0.8\linewidth,angle=0]{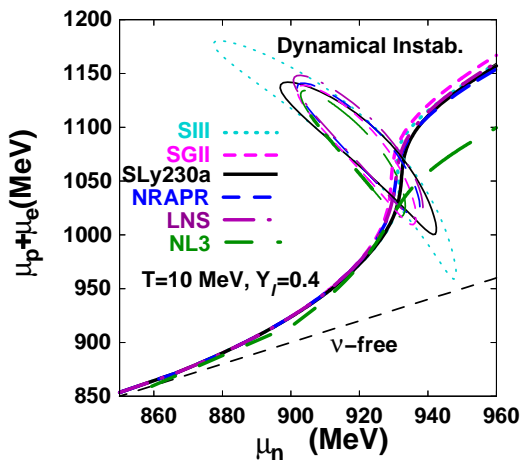}&
\end{tabular}
\end{center}
\caption{
(Color online)
Comparison between the $\beta$-equilibrium condition in homogeneous compact-star matter 
and the instability region in the chemical potential representation,
at finite temperature $T=10$~MeV. 
}
\label{FIG:spino-beq-mu}
\end{figure}

\begin{figure*}[thb]
\begin{tabular}{cc}
\includegraphics[width=1\linewidth,angle=0]{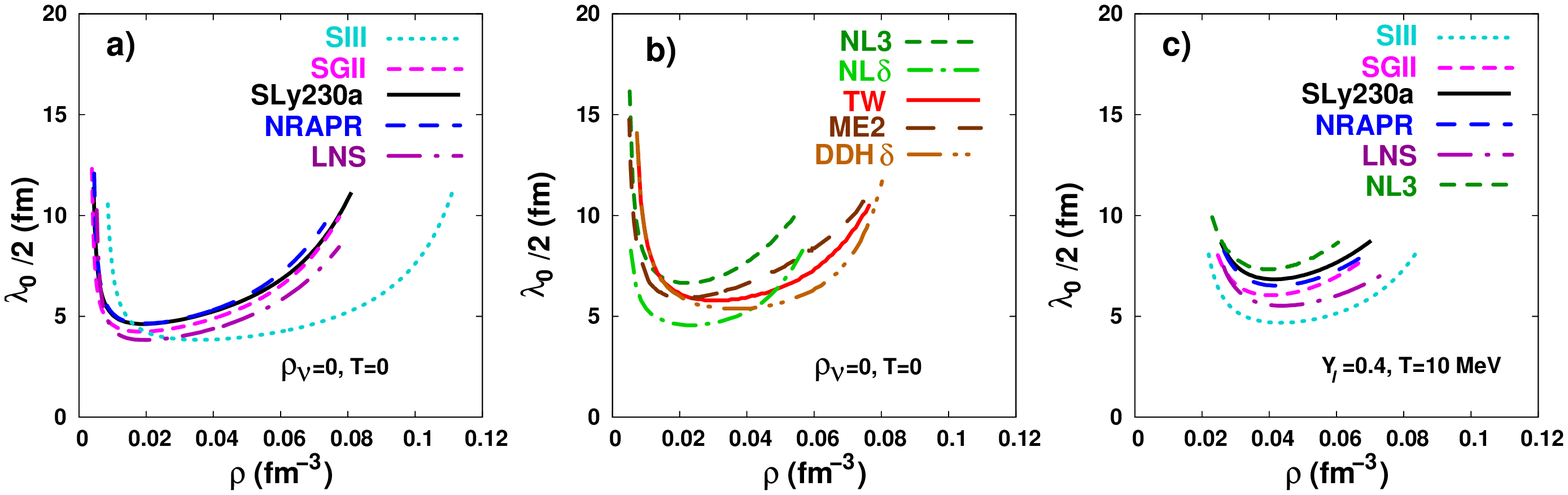}\\
\end{tabular}
\caption{
(Color online)
Half-wave-length of the most unstable mode inside the instability region.
At each density, the proton fraction is given 
by the $\beta$ equilibrium condition for homogeneous matter.
Graphs a) and b) : neutrino-free matter at $T=0$, respectively for Skyrme and relativistic models.
Graph c) : matter with trapped neutrinos (fixed lepton fraction $Y_l=0.4$) at $T=10$ MeV,
for Skyrme models and NL3.
}
\label{FIG:cluster}
\end{figure*}


Let us now comment on the supernova context.
A well-known issue in type II supernova simulation is the difficulty to obtain
the ejection of the outer layers of the collapsing star~\cite{Buras-PRL90}.
An additional mechanism is needed to produce a shock revival after the first bounce,
and it is generally assumed that neutrino transport is a crucial factor 
in the explosion dynamics.
It has been proposed that the liquid-gas instabilities in supernova matter could play an
important role, since it affects the transport properties of the neutrinos~\cite{Margueron-PRC70}.
We are then interested in comparing the dynamic spinodal region
to the $\beta$-equilibrium condition, in the situation that can be found in a type-II supernova core:
high temperature and possibly neutrino trapping.
In Ref.~\cite{CD-A3}, it was shown that an interplay occurs between neutrino trapping and cluster formation at temperatures of several MeV. Indeed, neutrino presence leads to more symmetric matter at $\beta$-equilibrium, reaching the instability region even though it is reduced by temperature;
reciprocally, cluster formation is known to favor neutrino trapping.
This feature is illustrated here on Fig.~\ref{FIG:spino-beq-mu}.
We show the dynamic spinodals at $T=10$~MeV in the chemical-potential representation
($\mu_n,\mu_p+\mu_e$), where the $\beta$-equilibrium is a model-independent straight line,
the diagonal $\mu_n=\mu_p+\mu_e$.
At this temperature, none of the instability regions reaches this line;
however, all are crossed by the $\beta$-equilibrium condition
for a constant lepton fraction $Y_l=0.4$.
The distance between each spinodal and the diagonal
measures the neutrino chemical potential which is needed to reach the instability region.
This is related to the $\mu_3$ extension of the spinodals,
which gives an example of distinction between models 
having high or low values of $a_{\s}$ at subsaturation density: 
disregarding SIII, the smallest neutrino trapping is needed for SLy230a, 
and the highest is for NL3.

We finally discuss the typical cluster size obtained according to the spinodal-instability properties,
in stellar-matter condition.
As in section~\ref{SUBSEC:dyn-spino}, 
we define it as the half-wavelenght of the most unstable mode, $\lambda_{0}/2$.
This quantity is given on Fig.~\ref{FIG:cluster} as a function of the density, 
in two different conditions:
cold, neutrino-free matter (for all models) 
and hot matter with trapped neutrinos (for Skyrme models and NL3). 
For each density, we estimate the proton fraction 
according to the $\beta$-equilibrium condition in homogeneous matter
\footnote
{
We should note that the $\beta$-equilibrium condition, established for homogeneous matter, 
is only intended here to give an estimation of the relevant proton fraction.
Indeed, since we are inside the instability region, the homogeneous matter is fastly decomposed into clusters, much before it can reach the $\beta$-equilibrium associated with this density.
On the other hand, a calculation of $\beta$-equilibrium in the clusterized medium
does not enter our study, which addresses the dynamic instability properties of homogeneous matter.
}.
Due to the weak dependence of $\lambda_{0}$ on the system asymmetry,
the curves we obtain essentially reflect the features
we had obtained earlier for $\lambda_{0}$ at $Y_{p}=0.3$ (Fig.~\ref{most_inst}).
At $T=0$, we  notice that the high asymmetry 
leads to a global increase of the cluster sizes by $\sim 1$~fm;
a similar behavior was obtained in Refs.~\cite{umodes06,CD-A3}.
The effect of temperature is to 
reduce the instability region and increase the cluster size
(typically by $\sim (2-3)$~fm for $T=10$~MeV).
We finally remark that the range of cluster sizes we obtain in this spinodal-scenario approach
is in reasonable agreement with pasta-phases calculations
performed in RMF framework~\cite{maruyama05,avancini08}.


\section{\protect\smallskip Conclusions}

The present work provides a direct comparison of Skyrme and relativistic models predictions
for nuclear and compact-star matter properties,
involving the bulk equation of state and the finite-size liquid-gas instabilities.
In our comparison, many similarities are found, and some differences are pointed out.

As expected, the largest differences are obtained for the isovector properties.
The relativistic models with constant couplings have a very hard symmetry energy.
On the opposite, the older Skyrme forces 
predict a decrease of the symmetry energy at high density. 
Between these two extremes, modern Skyrme forces and DDH models 
present similar behaviors.
However, we can notice the specificities of DDH$\delta$ and SLy230a
in the high density region:
both present a stiffer $a_{\s}$ evolution, and atypic rates $\mu_3/\mu_3^{\rm{para}}$
(for DDH$\delta$, this is linked to the inclusion of the $\delta$ meson).

Concerning the thermodynamic spinodal region,
we have verified that its isoscalar-density extension $\rho_{\s}$
is correlated to the saturation density $\rho_0$.
At zero temperature, the spinodal contour reaches very high asymmetry for all models;
using the $(\rho,\mu_3)$ representation, 
we have obtained different $\mu_3$ extensions, 
reflecting the low density values of the symmetry energy.
Relativistic models tend to yield smaller spinodal regions, both in $\rho$ and $\mu_3$ directions.
The thermodynamic instability direction leads to the usual isospin distillation for all models;
however  for larger densities this effect becomes stronger 
for RMF models with constant couplings and presents a reduction for
 Skyrme and DDH models.
Some very recent DBH results seem to confirm this trend, 
although with a smaller reduction~\cite{isaac08}.

The dynamic finite-size instabilities leading to matter clusterization 
have been addressed in the Vlasov formalism.
The wavelength associated with the most unstable mode
gives an estimation of the cluster size issuing from a spinodal decomposition:
it was shown that Skyrme parametrizations show larger growth rates
and favor fluctuations of shorter wavelengths.
The favored wavelengths reflect the finite range behaviour of the force.
In the Skyrme parametrizations, only $q^2$ terms are included.
With relativistic models, the finite range is described by the exchange of mesons. 
For all models, the dynamic distillation effect is reduced with respect to the bulk one;
this reduction appears stronger for relativistic models including the $\delta$ meson.

For the study of compact-star matter properties, 
we have considered $\beta$-equilibrium under two different conditions:
for neutrino-free matter, and for matter with trapped neutrinos according to a fixed lepton fraction.
For neutrino-free matter, the proton fraction at $\beta$-equilibrium 
is very sensitive to the symmetry energy $a_{\s}(\rho)$.
The RMF models with constant couplings thus predict very high proton fractions, 
allowing the direct URCA process already at quite low densities.
However, it was observed that the proton-fraction 
is not uniquely determined by $a_{\s}(\rho)$,
but also depends on the parabolic behavior of the isovector EOS:
see SLy230a and DDH$\delta$, leading to  proton fractions
lower than  other models with similar symmetry energy.
The situation is different for matter with trapped neutrinos: 
fixing a lepton fraction $Y_l=0.4$,
all models (except SIII) predict a nearly constant proton fraction $\sim 0.3-0.35$.

We have finally discussed clusterization of stellar matter in two different contexts:
at zero temperature, where it is related to the extension of the neutron-star crust,
and at finite temperature, where it should influence supernova dynamics.
In the first case, we give a lower estimation ($\rho_{\rm{cross,in}}$) 
of the transition density at the inner edge of the crust.
Our results are in reasonable agreement with values obtained within pasta-phase calculations. 
Modern Skyrme parametrizations and DDH models give similar results, 
$\rho_{\rm{cross,in}}\sim 0.75$~fm$^{-3}$.

Stellar matter at finite temperature is addressed only with Skyrme models and NL3.
Finite-$T$ calculations still have to be performed for the relativistic models 
with density-dependent couplings,
but we do not expect those results will affect our present conclusions.
For all the models we show, 
neutrino trapping is needed to reach the instability region at $T=10$~MeV.
The required trapping rate is higher 
for models with low symmetry energy at subsaturation density, such as NL3.

Globally, we can say that the isovector EOS shows quite large quantitative differences
even between modern forces:
new data are still needed to better constrain the neutron-rich matter properties.
We have shown the consequences of the stiff symmetry energy of the RMF models:stronger distillation, smaller $\mu_3$-extension of the spinodal, and larger proton-fractions at high density, allowing the URCA process.
The finite-range of the force is also important:
its different behavior between Skyrme and relativistic models causes 
the difference in the predictions of typical cluster size.

A similar work comparing phenomenological models with results from other approaches 
such as Brueckner-Hartree-Fock should be performed.

\section*{ACKNOWLEDGMENTS}

CD thanks the Coimbra University group for warm hospitality.
This work was partially supported by POCI2010 and FCT (Portugal) under the projects 
POCI/FP/63918/2005 and by POCI/FP/81923/2007.


\end{document}